\documentclass[lineno]{aa}  
\usepackage{graphicx}
\usepackage{txfonts}
\usepackage{subfigure}
\usepackage{ulem}

\usepackage[breaklinks=true, colorlinks=true, urlcolor=magenta, citecolor=blue, linkcolor=blue]{hyperref}

\newcommand{\orb}{${\Omega}$\,} 
\newcommand{\spin}{${\omega}$\,} 
\newcommand{\be}{${\omega-\Omega}$\,} 
\newcommand{\porb}{P$_{\Omega}$\,} 
\newcommand{\psp}{P$_{\omega}$\,} 
\newcommand{\pbe}{P$_{\omega-\Omega}$\,} 

\begin{document}

\title{ Precise Timing Analysis of Four Magnetic Cataclysmic Variables with TESS}
    \author{Srinivas M Rao\inst{\ref{ARIES} ,\ref{MJPRU}}\thanks{srinivas@aries.res.in} \and Jeewan C Pandey\inst{\ref{ARIES}}
   \and Nikita Rawat\inst{\ref{ARIES},\ref{SAAO}}
   \and Arti Joshi \inst{\ref{PUC}}
   \and Ajay Kumar Singh \inst{\ref{BAR}}        
       }
       
    \institute{
    Aryabhatta Research Institute of Observational Sciences, Manora Peak, Nainital 263001, India\label{ARIES}
    \and
    Mahatma Jyotiba Phule Rohilkhand University, Bareilly 243006, India \label{MJPRU}
    \and
    South African Astronomical Observatory, PO Box 9, Observatory, 7935, Cape Town, South Africa \label{SAAO}
    \and
    Institute of Astrophysics, Pontificia Universidad Católica de Chile, Av. Vicuña MacKenna 4860, 7820436 Santiago, Chile \label{PUC}
    \and
    Department of Applied Physics/Physics, Bareilly College, Bareilly-243001, India \label{BAR}
    }
    
   \date{}

 
  \abstract
   {We analysed high time-resolution optical photometric data from the Transiting Exoplanet Survey Satellite (TESS) to study the timing behaviour of four intermediate polar-like objects, namely, V1460 Her, 1RXS J045707.4+452751, Swift J0958.0-4208, and V842 Cen. In the case of V1460 Her, we refined the measurement of its orbital period. Long-term observations suggest a gradual decrease in the orbital period of V1460 Her, and the stable light curve during the TESS observations indicates its quiescent state. We detected a beat period of 1290.6 $\pm$ 0.5 s for the first time for the source 1RXS J045707.4+452751, suggesting a possible disc-overflow accretion scenario. For the sources Swift J0958.0-4208 and V842 Cen, we determined periods 6.11 $\pm$ 0.02 h and 3.555 $\pm$ 0.005 h, respectively, which can be provisionally suggested to be orbital periods. These findings provide valuable insights into the accretion processes and long-term evolution of these intriguing binary systems.
  }

    \authorrunning{Rao et~al.}
   \maketitle

\section{Introduction}
\label{sec:intro}
Cataclysmic variables (CVs) are semi-detected interacting binary systems characterised by the dynamic interaction between a white dwarf (WD), referred to as the primary, and a late-type main-sequence star, known as the secondary. The secondary overflows its Roche lobe, leading to material accretion onto the WD \citep{Warner1995}. These systems are categorised into two distinct groups based on the strength of the WD's magnetic field: non-magnetic CVS (NMCVS) and magnetic CVS (MCVS). NMCVs are identified by their relatively weak magnetic fields, typically measuring less than 0.1 million Gauss (MG), which results in the magnetic field having minimal influence on the accretion process from the secondary star to the WD. In contrast to NMCVS, MCVS are characterised by significantly stronger magnetic fields and have two subclasses: polars and intermediate polars (IPS). Polars are unique for their exceptionally powerful magnetic fields, exceeding 10 MG \citep{Cropper1990}. On the other hand, IPS exhibit magnetic field strength typically ranging from 1 to 10 MG \citep{Patterson1994}. 

In NMCVS, the material transferred from the secondary forms a disc around the WD and then falls on its surface. On the other hand, in polars, the strong magnetic field inhibits the formation of the disc altogether, and the material is directly channelled to the poles of the WD. In addition to the magnetic field, other factors, such as mass accretion rate and binary separation, play an important role in governing the accretion mechanism in IPS. Three distinct accretion mechanisms, namely disc-fed, stream-fed, and disc-overflow, are generally observed in IPS. In the disc-fed mechanism, the accretion process occurs from a Keplerian disc, and this disc is disrupted at a specific point known as the magnetospheric radius of the WD, and then the material follows along the magnetic field lines \citep[e.g.][]{Hellier1989a}. Within the stream-fed mechanism, the material is directly accreted onto the surface of the WD by following the magnetic field lines \citep{Rosen1988}. However, in the disc-overflow mechanism, material overflows from the accretion disc and collides with the magnetosphere. Consequently, both modes of accretion, stream-fed and disc-fed, operate concurrently as described in \citet{Hellier1989b}. The power at the spin (\spin) and the beat (\be) frequencies in the power spectra gives us an idea about the probable accretion geometry of the system \citep{Hellier1991, Hellier1993}. Near the WD surface, the matter falls freely on the surface and approaches supersonic velocities, which causes shocks to form above the WD surface. The material decelerates before settling onto the surface and emits cyclotron (optical/infrared) and bremsstrahlung (X-ray) radiation \citep{Aizu1973}. The optical emission range is generated from the reprocessing of the X-rays from different regions of the WD, such as from the surface of the WD or the accretion disc surrounding it.

In this paper, we have performed a detailed timing analysis of four IP-like sources using the short cadence long baseline \textit {TESS} observations. These  are V1460 Her, 1RXS J045707.4+452751, Swift J0958.0-4208, and V842 Cen. We were interested in finding new periods in these sources and also confirming the periods obtained from the earlier studies. These periods are essential for confirming the nature of these sources and also their accretion geometry. A review of each source is given in the following paragraphs. 

\begin{table*}
    \centering
    \caption{Observation log of V1460 Her, J045707, J0958, and V842 Cen.}
    \begin{tabular}{l l c c c}
    \hline
     Source   & Sector & Start Time & End Time & Total observing days\\
    \hline
    V1460 Her & ~~24 & 2020-04-16 07:06:44 & 2020-05-12 18:46:58 & 26.49 \\
              & ~~25 & 2020-05-14 03:12:57 & 2020-06-08 19:22:26 & 25.67 \\
              & ~~50 & 2022-03-26 18:31:36 & 2022-04-22 00:16:23 & 26.24 \\
              & ~~51 & 2022-04-23 11:00:24 & 2022-05-18 00:52:28 & 24.58 \\
              & ~~52 & 2022-05-19 03:16:28 & 2022-06-12 13:51:53 & 24.44 \\
              & ~~77 & 2024-03-25 23:41:08 & 2024-04-23 01:13:18 & 28.06 \\
              & ~~79 & 2024-05-22 00:57:18 & 2024-06-18 04:12:32 & 26.14 \\
    J045707   & ~~59 & 2022-11-26 18:30:00 & 2022-12-23 04:36:00 & 27.14\\
    J0958     & ~~9  & 2019-03-01 02:14:44 & 2019-03-25 23:28:48 & 24.88\\
    V842 Cen$^\bigstar$ & ~~38 & 2021-04-29 08:35:10 & 2021-05-26 01:19:28 & 26.70\\
    \hline
    \end{tabular}
    ~~\\
    $^\bigstar$ The data is available for the 20s cadence too.
    \label{tab:obslog}
\end{table*}

\begin{table}
\addtolength{\tabcolsep}{-5pt}
    \centering
    \caption{Periods obtained corresponding to the dominant peaks of the LS power spectra of V1460 Her, J045707, J0958, and V842 Cen.}
    \label{tab:ps}
{\tiny
    \begin{tabular}{l c c c c c c c c c}
    \hline
     Source   & Sector & Cadence & $P_\Omega$ & $P_\omega$& $P_{\omega-\Omega}$ \\
     \hline
     V1460 Her & 24 & 120 & $4.985 \pm 0.009$* & - & - \\
              & 25 & 120 & $4.988 \pm 0.010$* & - & - \\
              & 50 & 120 & $4.987 \pm 0.009$* & - & - \\
              & 51 & 120 & $4.988 \pm 0.011$* & - & - \\
              & 52 & 120 & $4.981 \pm 0.011$* & - & - \\
              & 77 & 120 & $4.989 \pm 0.009$* & - & - \\
              & 79 & 120 & $4.990 \pm 0.009$* & - & - \\
              & Com$^\dagger$ & & $4.9885 \pm 0.0002$* & - & - \\
    J045707 & 59 & 120 & $6.09 \pm 0.02$ & $1218.7 \pm 0.2$ & $1290.5 \pm 0.2$ \\
    J0958 & 9 & 120 & $6.11 \pm 0.02$ & $296.12 \pm 0.01$ \\
    V842 Cen & 38 & 120 & $3.555 \pm 0.005$ & - & - \\
             & 38 & 20 & $3.555 \pm 0.005$ & - & - \\
    \hline
    \end{tabular}\\
{$\dagger$ Represents the periods derived from the power spectra of combined \textit{TESS} observations of all sectors for V1460 Her. * Harmonics of the \orb frequency for all sectors' and combined power spectra were also found.} 
}
\end{table}

V1460 Her was initially identified as a contact W UMa eclipsing binary \citep{Pollocco2006} with a period of 0.208 d \citep{Lohr2013} and an overluminous K5-type donor star. However, \cite{Drake2016} reported an outburst and suggested it as an unusual CV.  Based on the spectroscopic observations, \cite{Scaringi2016} found dwarf-nova outbursts in this system. \cite{Kjurkchieva2017} derived an orbital inclination of $88^\circ \pm 3^\circ$ for V1460 Her. It was confirmed as an IP after identifying a spin period of WD (\psp) of 38.875 s \citep{Ashley2020,Pelisoli2021}. 

1RXS J045707.4+452751 (hereafter J045707) was identified as X-ray source by \cite{Kaplan2006} using the \textit{Chandra} observations. It was first classified as an NMCV \citep{Masetti2010}, but later, from the optical spectra, it was proposed to be an MCV or a high accretion rate nova-like source \citep{Masetti2012}. \cite{Thorstensen2013} detected a \psp of $1218.7\pm0.5$ s and classified it as a magnetic CV. They have also obtained a probable orbital period (\porb) of either 4.8 h or 6.2 h using optical spectroscopy observations. Using the X-ray data obtained from \textit{XMM-Newton} \cite{Bernardini2015} found the \psp as $1222.6\pm2.7$ s and the mass of the WD ($M_{WD}$) as $1.12 \pm 0.06 M_\odot$. Later \cite{Suleimanov2019} estimated the mass of WD as $0.87 \pm 0.14 M_\odot$. Using \textit{TESS} data, \cite{2024AJ....168..121B} has recently derived an orbital period of $6.091\pm0.007$ h.

Swift J0958.8-4280 (hereafter J0958) was identified as an MCV by \cite{Masetti2013} from the \textit{INTEGRAL} observations. \cite{Bernardini2017} found  \psp as $296.22 \pm 0.05$ s using X-ray data from XMM-Newton. They suggested that more observations are required for further classification. 

V842 Cen (=Nova Centauri 1986) was discovered by \cite{McNaught1986} on November 22.7 UT when it was in nova state. It took 48 d to decline its brightness by 3 mag, suggesting that it was a moderately fast nova \citep{Sekiguchi1989}. High-speed photometry by \cite{Woudt2003} showed that the system was still active and continuously flaring on timescales of 5 min but showed no orbital modulation. Hence, they concluded that the system was seen at a low inclination. However, \cite{Woudt2009} again did optical photometry observations in 2008 and obtained a pulsation period of 56.825 s, which they attributed as \psp of the WD. Based on the presence of sidebands, they inferred an \porb of 3.94 h and classified it as an IP. \cite{Luna2012} detected a period of 3.51 h, but they could not detect the 56.825 s  periodicity using X-ray data from \textit{XMM-Newton} data. Optical light curve analysis by \cite{Sion2013} revealed a 56.5 s periodicity, but this periodicity was not seen in the UV light curve created from the HST COS spectroscopic observations.  Later \cite{Warner2015} suggested a period of 113.6 s, which is twice that of the \psp earlier observed.

The paper is organized as follows. In section \ref{sec:obs}, we describe observations and data. Section \ref{sec:ana} contains the analysis and the results. Finally, we present the discussion and conclusions in sections \ref{sec:dis} and \ref{sec:conc}, respectively.

\begin{figure*}[!ht]
    \centering
    \vspace{-0.2cm}
    \subfigure[Long-term light curve of V1460 Her]{\includegraphics[width=0.88\textwidth]{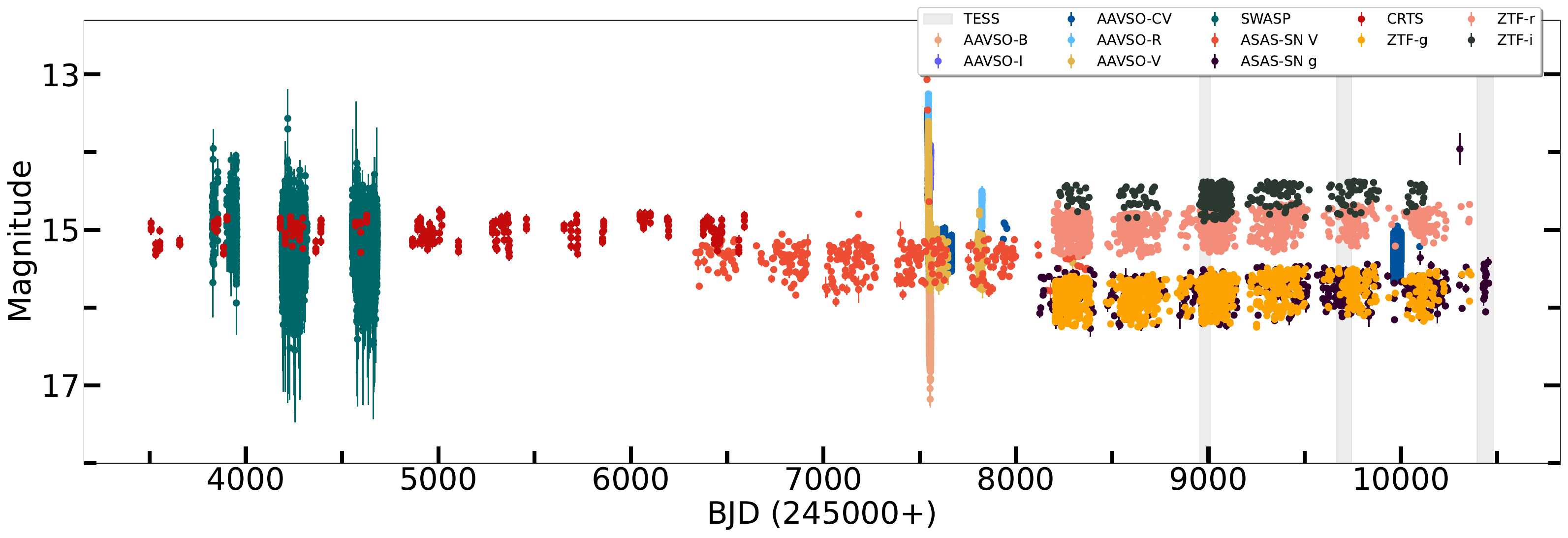}\label{fig:comblc_V1460}}
    \vspace{-0.3cm}
    \subfigure[\textit{TESS} light curve of two consecutive days of  sector 24]{\includegraphics[width=0.44\textwidth]{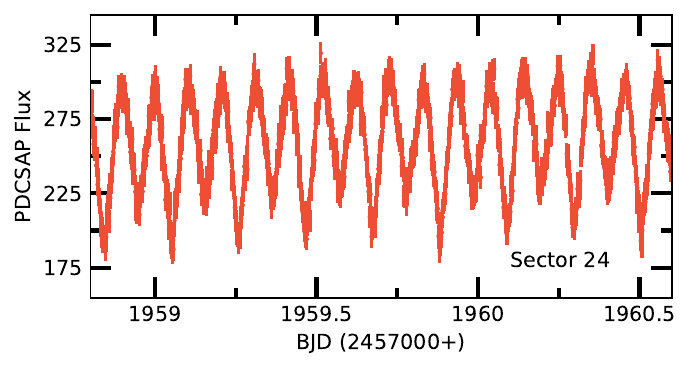}\label{fig:TESSlc_V1460}}
    \subfigure[AAVSO-CV light curve of V1460 Her]{\includegraphics[width=0.44\textwidth]{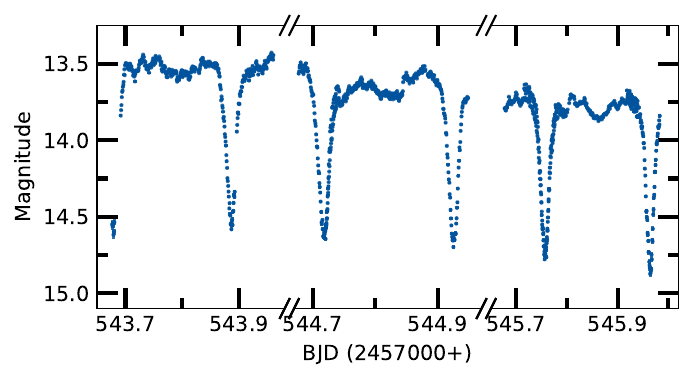}\label{fig:aavsolc_V1460}}
    \subfigure[Power Spectra]{\includegraphics[width=0.88\textwidth]{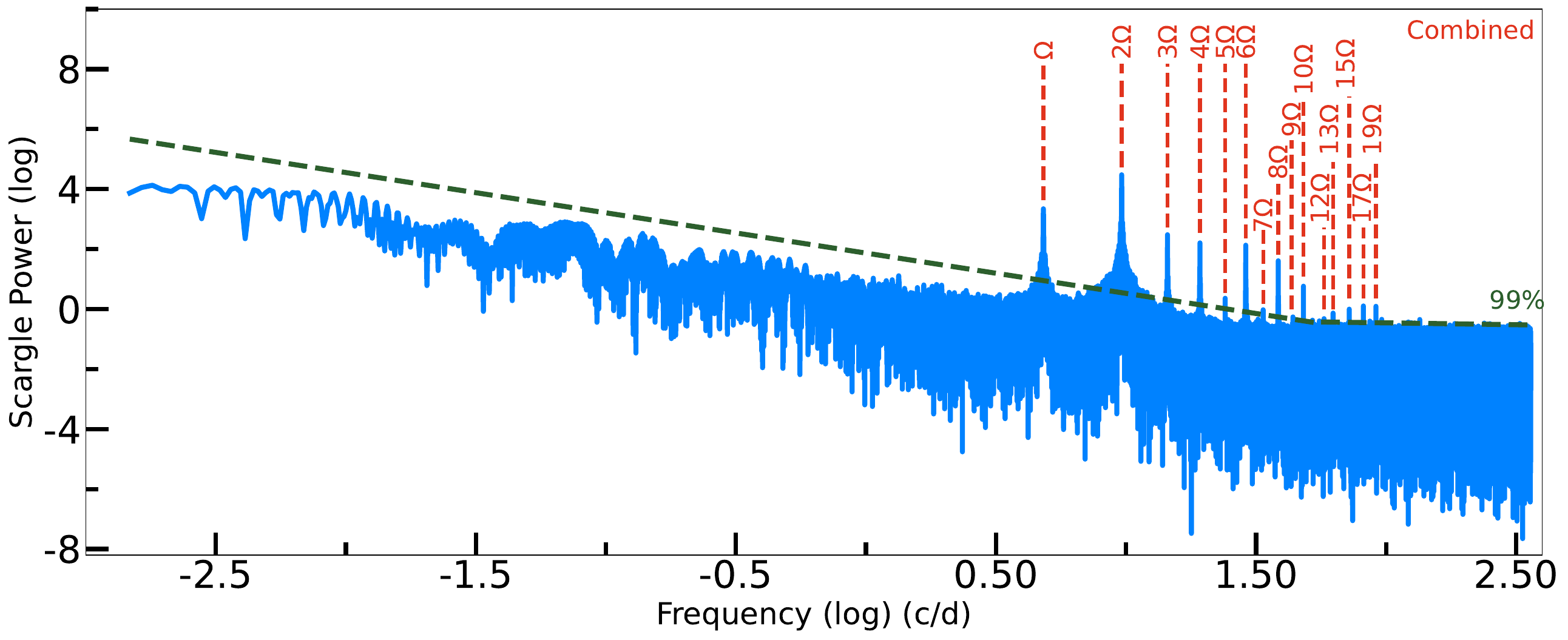}\label{fig:ps_V1460}}
    \vspace{-0.0cm}
    \caption{(a) Long-term light curve of V1460 Her (see text for detail). (b)\textit{TESS} light curve of approximately two consecutive days of observations of V1460 Her for sector 24. (c) AAVSO-CV light curve of V1460 Her of three consecutive days. (d) \textit {TESS} power spectra of all combined \textit{TESS} dataset of V1460 Her, where the identified frequencies are marked with red vertical dashed lines. The green dashed line represents the 99\% confidence level.}
\end{figure*}

\section{Observations and data}
\label{sec:obs}
We have used the data obtained from high cadence long-baseline \textit{TESS} observations. The \textit{TESS} \citep{Ricker2015} instrument consists of four wide-field CCD cameras that can image a region of the sky, measuring $24^\circ \times 96^\circ$. \textit{TESS} observations are broken up into sectors, each lasting two orbits, or about 27.4 days and conducts its downlink of data while at perigee. This results in a small gap in the data compared to the overall run length. \textit{TESS} bandpass extends from 600 nm to 1000 nm with an effective wavelength of 800 nm. The data of our sources was stored in the Mikulski Archive for Space Telescopes data\footnote{\url{https://mast.stsci.edu/portal/Mashup/Clients/Mast/Portal.html}} with unique identification numbers `TIC 193515431', `TIC 65820714', `TIC 35975843', and `TIC 411603422'  for V1460 Her, J045707, J0958 and V842 Cen, respectively. We have used the Pre-search Data Conditioned Simple Aperture Photometry (PDCSAP) flux, which is the Simple Aperture Photometry (SAP) flux values corrected for instrumental variations\footnote{See section 2.1 of the TESS archive manual at \url{https://outerspace.stsci.edu/display/TESS/2.1+Levels+of+data+processing}}. PDCSAP flux also corrects for flux captured from the nearby stars. Data taken during an anomalous event had quality flags greater than 0 in the fits file. We have considered the PDCSAP flux data for which the quality flag is marked as `0'. The log of observations is shown in Table \ref{tab:obslog}. The cadence for each source was 120 s; however, 20 s cadence data was also available for V842 Cen.

\begin{figure}[!h]
    \centering
    \includegraphics[width=\columnwidth]{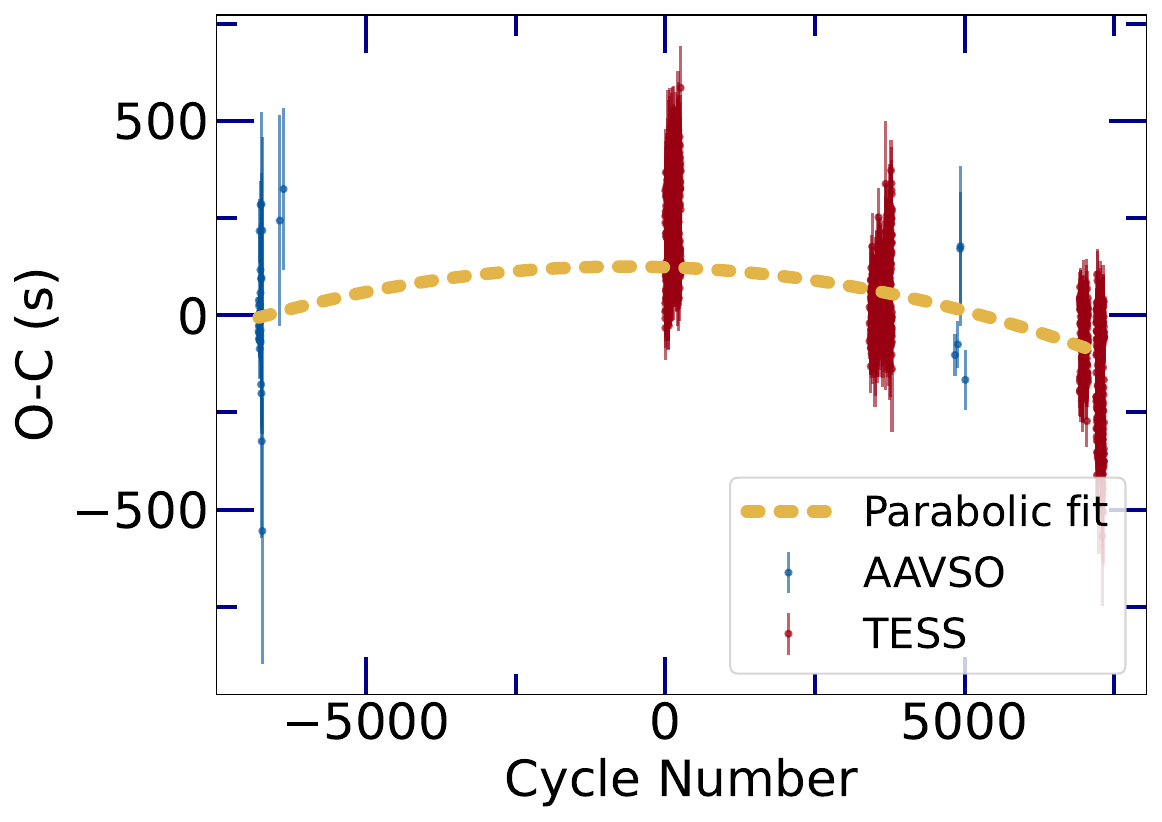}\label{fig:oc_V1460}
    \caption{Observed minus calculated (O-C) diagram for V1460 Her.}
\end{figure}

For V1460 Her we have also used the data from SuperWASP (Wide Angle Search for Planets; \citealt{Pollocco2006}), ASAS-SN (All Sky Automated Survey for Supernovae; \citealt{Shappee2014, Hart2023}), CRTS (Catalina Real-Time Transient Survey; \citealt{Drake2009}), AAVSO (American Association of Variable Star Observers), and ZTF (Zwicky Transient Facility; \citealt{Bellm2019}).

\section{Analysis and results}
\label{sec:ana}
\subsection{V1460 Her}
\subsubsection{Light curves and power spectra}
The long-term light curve of V1460 Her as observed from SuperWASP, ASAS-SN, CRTS, AAVSO, and ZTF is shown in Figure \ref{fig:comblc_V1460}. Timings of the \textit{TESS} observations are shown by a grey shaded region. V1460 Her was observed for seven sectors by \textit{TESS} (see Table \ref{tab:obslog}). Figure \ref{fig:TESSlc_V1460} shows the \textit{TESS} light curve of V1460 Her for sector 24 for approximately two consecutive days. The eclipsing nature of the light curves was clearly evident, with two eclipse-like profiles. The bigger dip in the flux was considered a primary eclipse, whereas the lower dip was considered a secondary eclipse. The time series data from all seven sectors were separately analysed using the Lomb-Scargle (LS) periodogram \citep{Lomb1976, Scargle1982} method to know the periodic behaviour of the light curve. The LS power spectral analysis was also performed on all the sectors' combined datasets. Figure \ref{fig:ps_V1460} shows the power spectra of the combined data set. The lower frequency limit was taken as 2/timespan of observations. The significance of peaks in power spectra was evaluated using the technique described by \cite{Vaughan2005}. Considering the impact of pink, white, and blue noises on \textit{TESS} power spectra \citep{Kalman2025}, we have accounted for these different noise components in our data. Since the data did not exhibit high-frequency noise, we excluded the blue noise component from our analysis. We detected only the significant orbital frequency (\orb) and its harmonics (marked in red dashed lines in Figure \ref{fig:ps_V1460}), which lie above the 99\% confidence level in the combined dataset and also in all the individual sectors. The 2\orb frequency was the most dominant frequency in the power spectra. In Table \ref{tab:ps}, we have provided the values of derived \porb in individual sectors. The previously reported \psp of 38.875 s by \cite{Ashley2020} exceeds the Nyquist limit of $\sim$360 cycle d$^{-1}$ (c/d) of current data; therefore, we were not able to detect the \psp.    

\subsubsection{Ephemeris and O-C analysis}
We have calculated eclipse midpoints by fitting the eclipses (both the primary and secondary eclipses) with a constant plus a Gaussian. The time at which we obtain the first primary eclipse of \textit{TESS} light curve was taken as the starting point of the orbital cycle. A total of  39 timings of minima from AAVSO and 1377 from \textit{TESS} were extracted. The data from CRTS, ASAS-SN, and ZTF were sparse, so we could not determine the exact minima from these light curves. These minimum timings were plotted against the cycle numbers. A linear fit between the cycle numbers and minima timings provides the following ephemeris :
\begin{equation}
    BJD = 2458955.93245(2) + 0.207852429(4) \times E \label{eqn:oc}
\end{equation}
 The numbers in parentheses are the errors on the corresponding parameters.   We obtained a refined orbital period of V1460 Her as $4.98845829 \pm 0.00000009$ h. We have also calculated the ephemeris after excluding the AAVSO data and found that the orbital period remained relatively consistent, with a difference of about 0.024 s; incorporating AAVSO data yielded a more precise ephemeris.  Further, folding the light curve with the ephemeris derived after excluding the AAVSO data did not alter the further analysis. The difference between observed and calculated (O-C) eclipse timings is shown in Figure \ref{fig:oc_V1460}, where a noticeable pattern is visible. By fitting a parabola between O-C and cycle numbers, the rate of the change in the \porb was estimated as $- (3.9\pm0.2) \times 10^{-10} ss^{-1}$. The negative sign indicates that the \porb is decreasing over the course of time.  

\subsubsection{Phase folded light curves}
\begin{figure*}
    \centering
    \subfigure[Day-wise orbital phase folded light curve]{\includegraphics[width=0.9\textwidth]{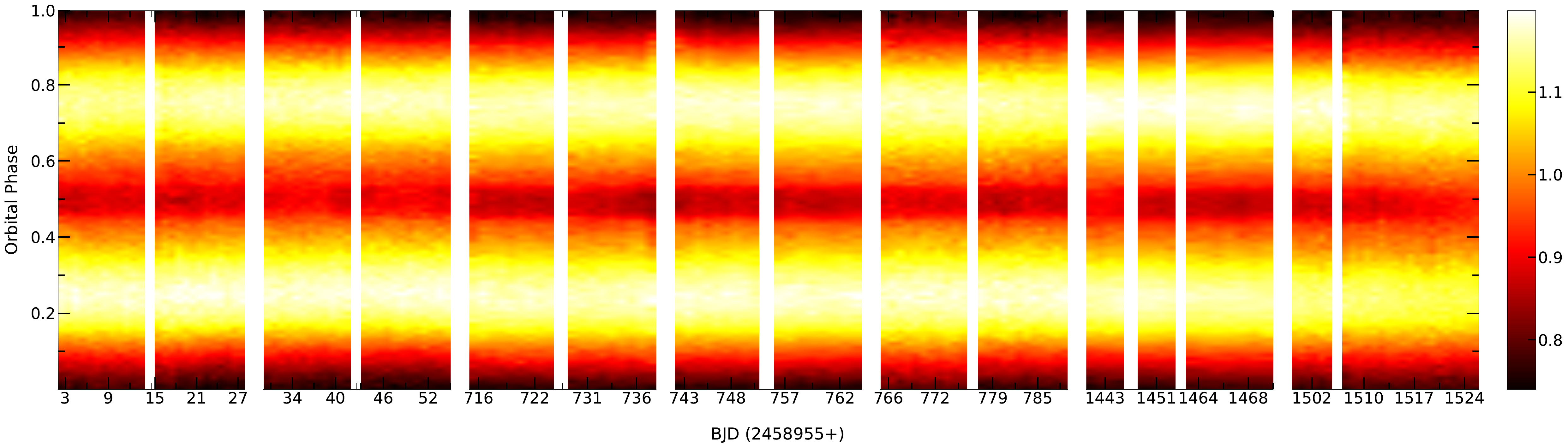}\label{fig:orbfold_V1460}}
    \subfigure[Phase folded light curves of V1460 Her at different epochs of observations]{\includegraphics[width=0.9\textwidth]{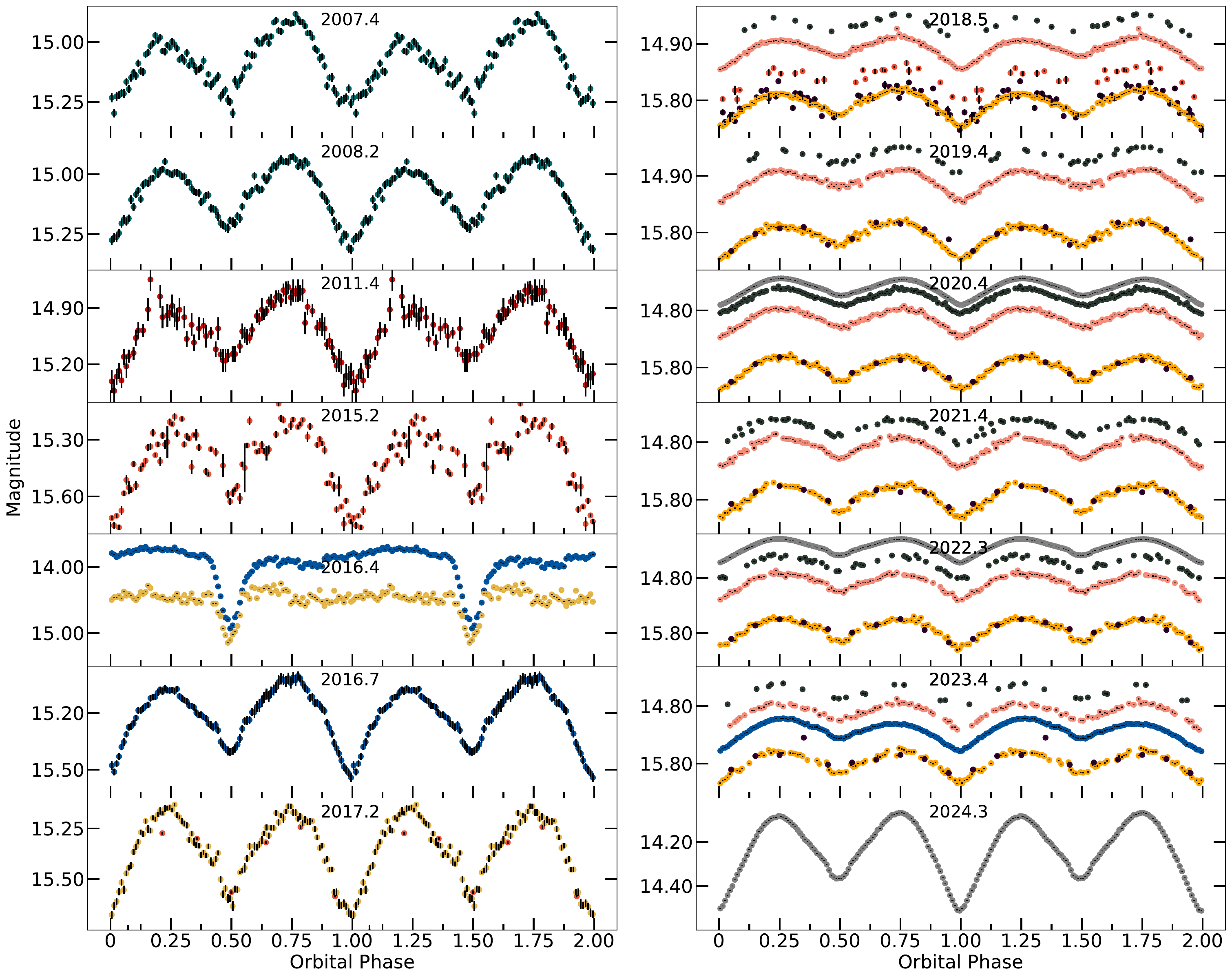}\label{fig:v1460_flc}}
    \caption{(a) Day-wise orbital phase folded light curve of V1460 Her of all sectors. The color bar on the right side indicates the normalised flux. (b) Orbital phase folded light curves of V1460 Her observed using the SuperWASP, CTRS, ASAS-SN, AAVSO, \textit{TESS}, and ZTF facilities at different epochs. The color schemes of the light curve are similar to  Figure  \ref{fig:comblc_V1460}. The epochs (in the year) are shown at the top of each panel. The phase bin size was chosen as 0.01.}
\end{figure*}

 The day-wise evolution of the orbital phased light curve for the \textit{TESS} data was also investigated and is shown in Figure \ref{fig:orbfold_V1460}. The white spaces in Figure \ref{fig:orbfold_V1460} show the gap in the data. The light curve was folded using the ephemeris as given in Equation \ref{eqn:oc}. Throughout all the sectors, we observed two maxima at phases 0.25 and 0.75. We did not find any significant change in the phases of the maxima in the day-wise folded light curves.

\begin{figure}
    \centering
    \subfigure[Representative light curve of J045707]{\includegraphics[width=\columnwidth]{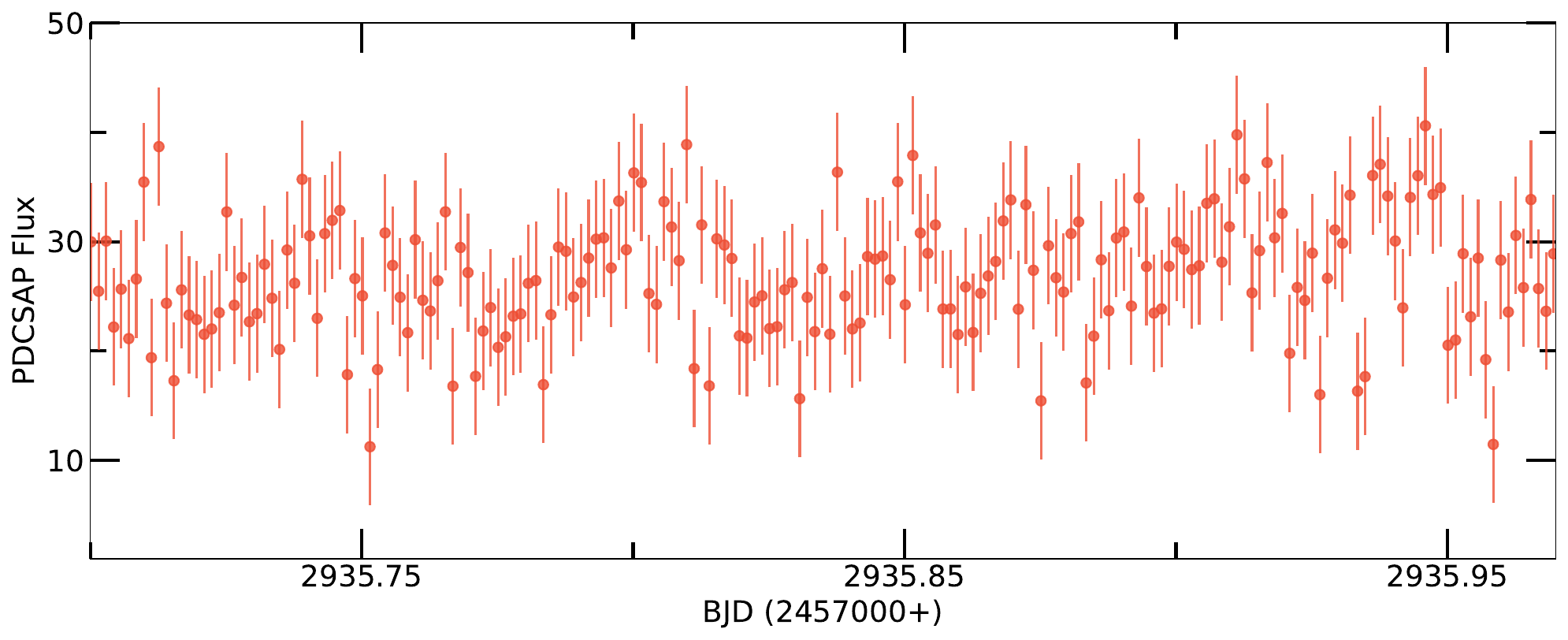}\label{fig:lc_J04}}
    \subfigure[Power spectra of \textit{TESS} observations of J045707]{\includegraphics[width=\columnwidth]{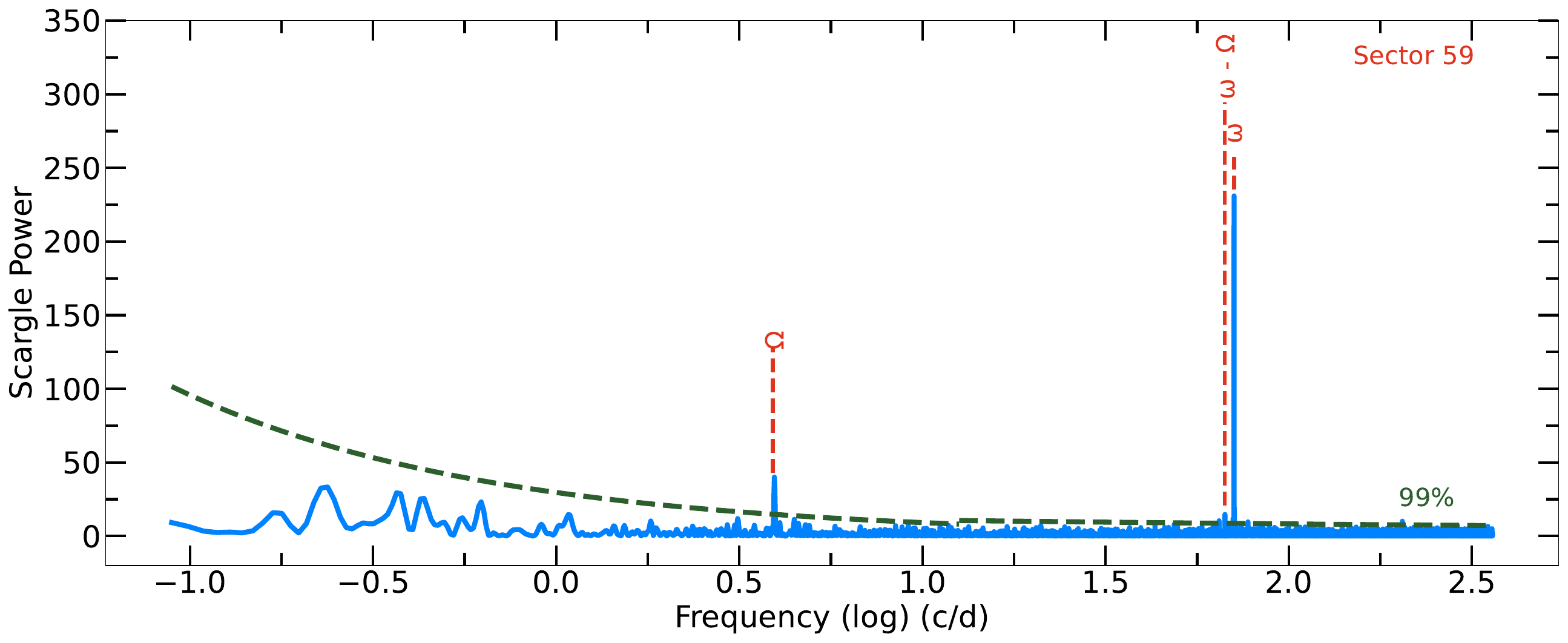}\label{fig:ps_J04}}
    \caption{(a) \textit{TESS} light curve of approximately one orbital cycle of J045707 from the observation of sector 59. (b) \textit {TESS} power spectra of J045707 for sector 59, where the identified frequencies are marked. The green horizontal dashed lines represent the 99\% confidence level. The $\omega - \Omega$ frequency is shown in the zoomed-in plot.}
    \label{fig:lc_ps_J04}
\end{figure}

\begin{figure}
    \centering
    \subfigure[Orbital phase folded light curve]{\includegraphics[width=\columnwidth]{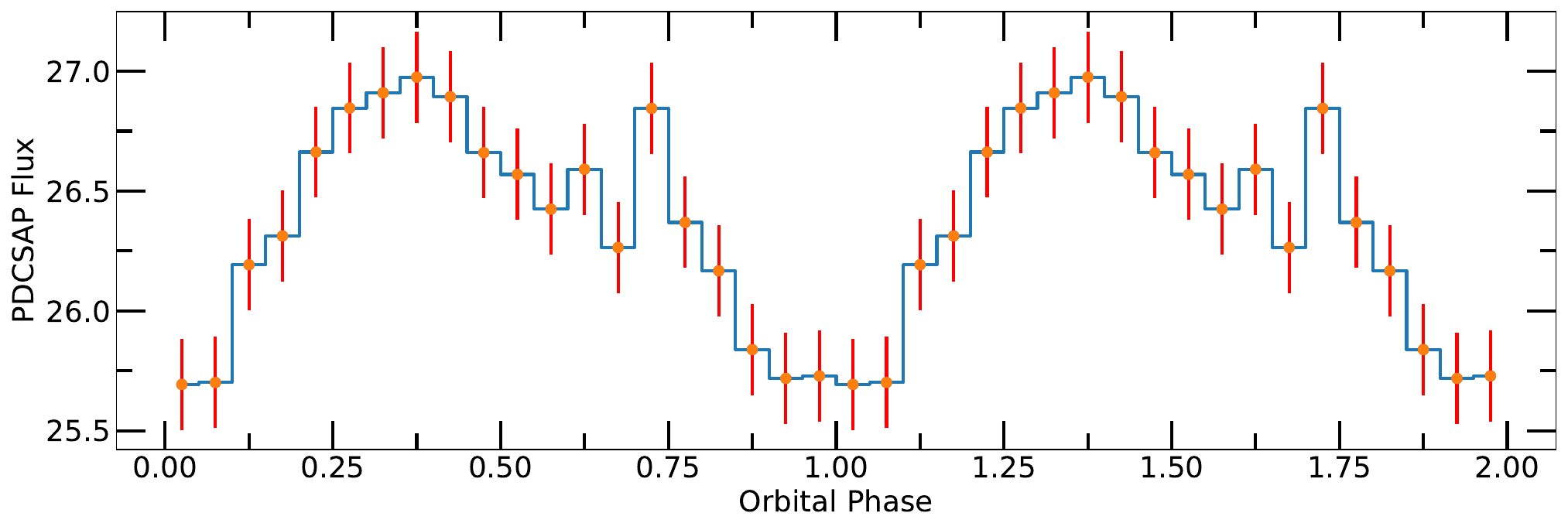}\label{fig:orbfold_J04}} 
    \subfigure[Spin phase folded light curve]{\includegraphics[width=\columnwidth]{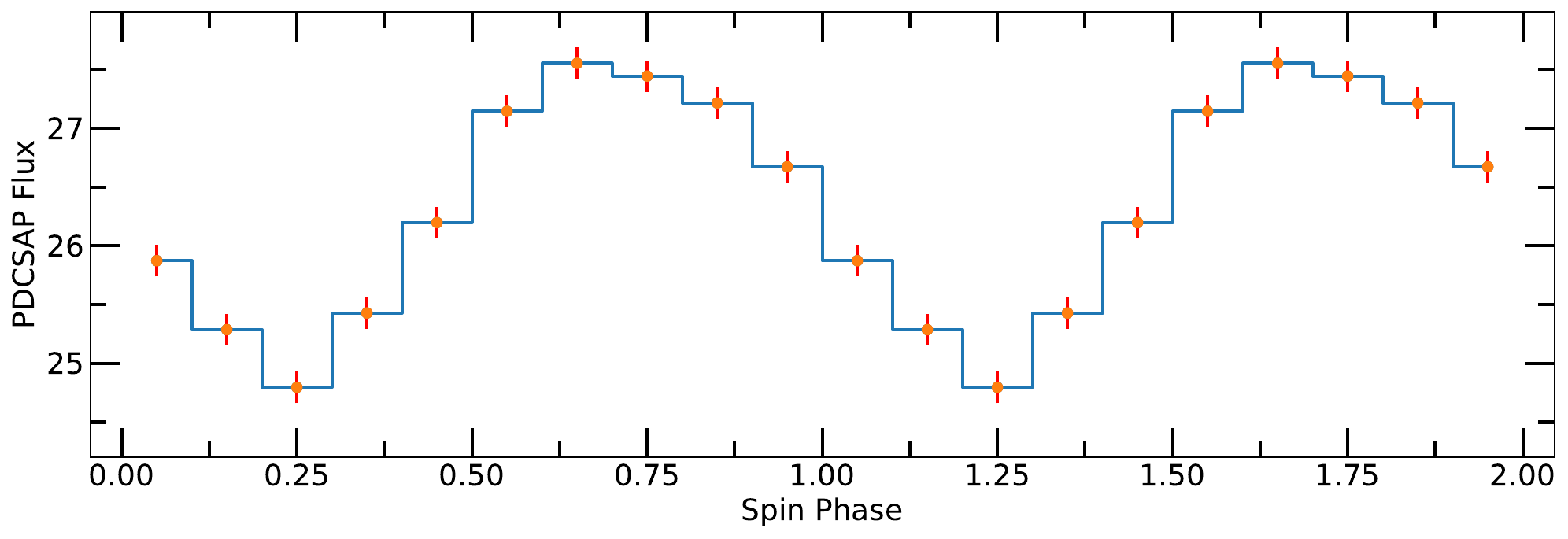}\label{fig:spfold_J04}} 
    \subfigure[Day-wise spin phase folded light curve]{\includegraphics[width=\columnwidth]{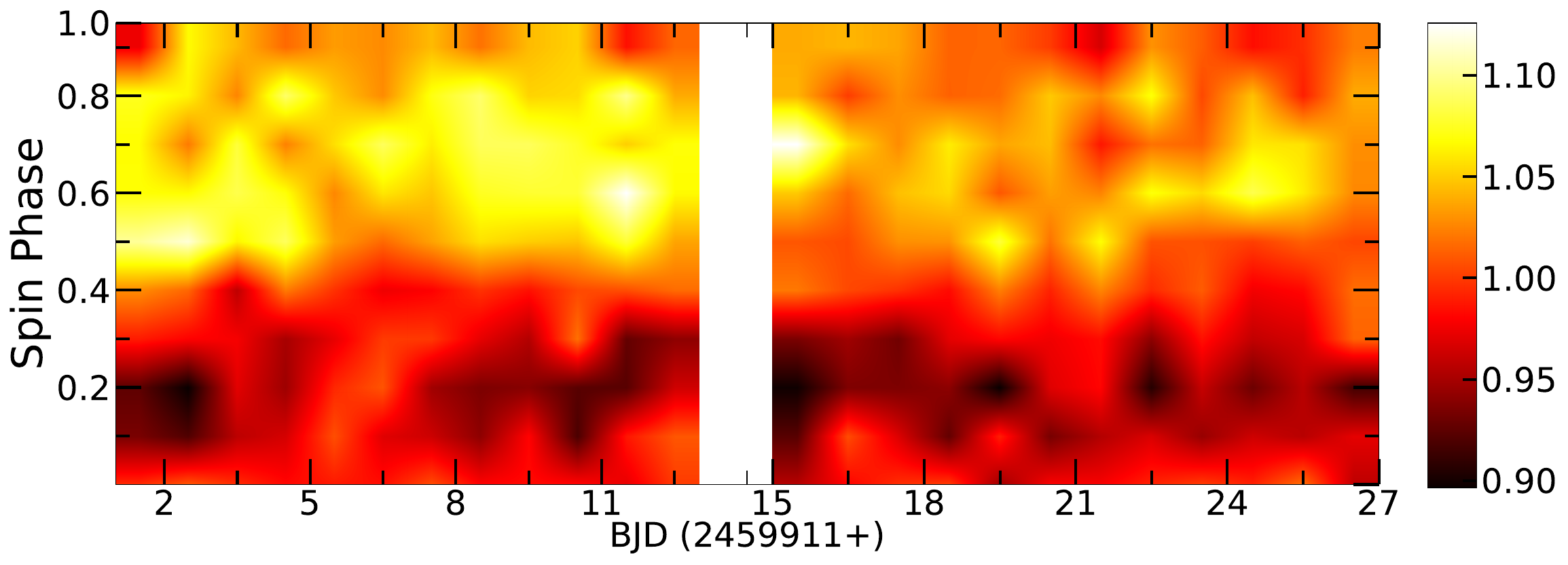}\label{fig:dspfold_J04}} 
    \caption{(a) Orbital and (b) Spin phase folded light curve of J045707 with a binning of 20. (c) Day-wise spin phase folded light curve of J045707 with a bin size of 0.05. The color bar represents the normalised flux}
\end{figure}

We have also compared the average phase folded light curves from different epoch of observations of V1460 Her, which are shown in Figure \ref{fig:v1460_flc}. All the light curves were folded using the ephemeris as given in Equation \ref{eqn:oc}. As noticed earlier by \cite{Kjurkchieva2017}, we have also found a different nature of the phased-light curve for epoch 2016.4, which might be in the outburst state as found by \cite{Drake2016}. The folded light curve pattern is observed to be distinct in both quiescence and outburst states. The quiescent profile exhibits double-peaked orbital modulations, with maxima occurring near phases 0.75 and 0.25 and shallow and deeper minima at phases 0.50 and 0, respectively. Both peaks appear to be of similar amplitude in some epochs, but in most epochs, they are of unequal amplitude. The presence of a double-humped feature is also evident from the power spectrum, where a strong significant peak is observed corresponding to the second harmonic of the orbital frequency. However, during the outburst at epoch 2016.4, a broader minimum appears, filled with emission and exhibiting stable brightness.

\begin{figure}
    \centering
    \subfigure[Representative light curve of J0958]{\includegraphics[width=\columnwidth]{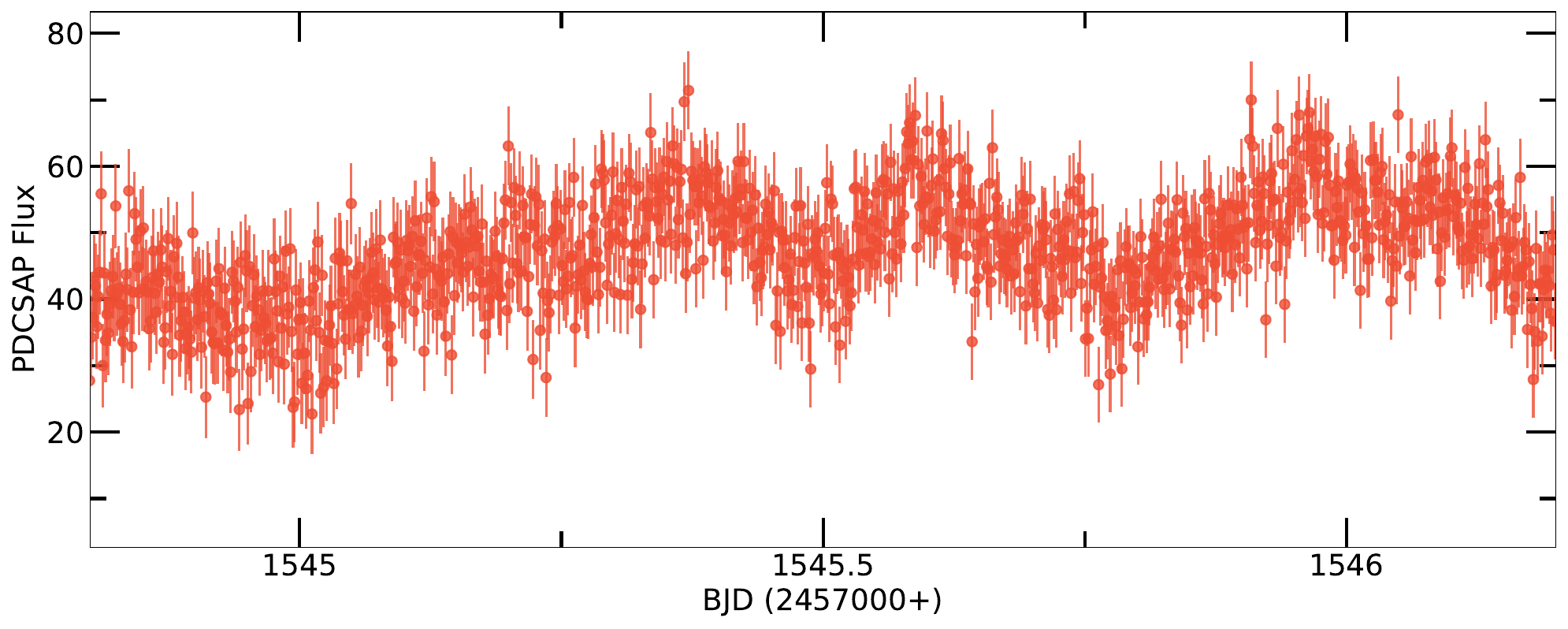}\label{fig:lc_J09}}
    \subfigure[Power spectra]{\includegraphics  [width=\columnwidth]{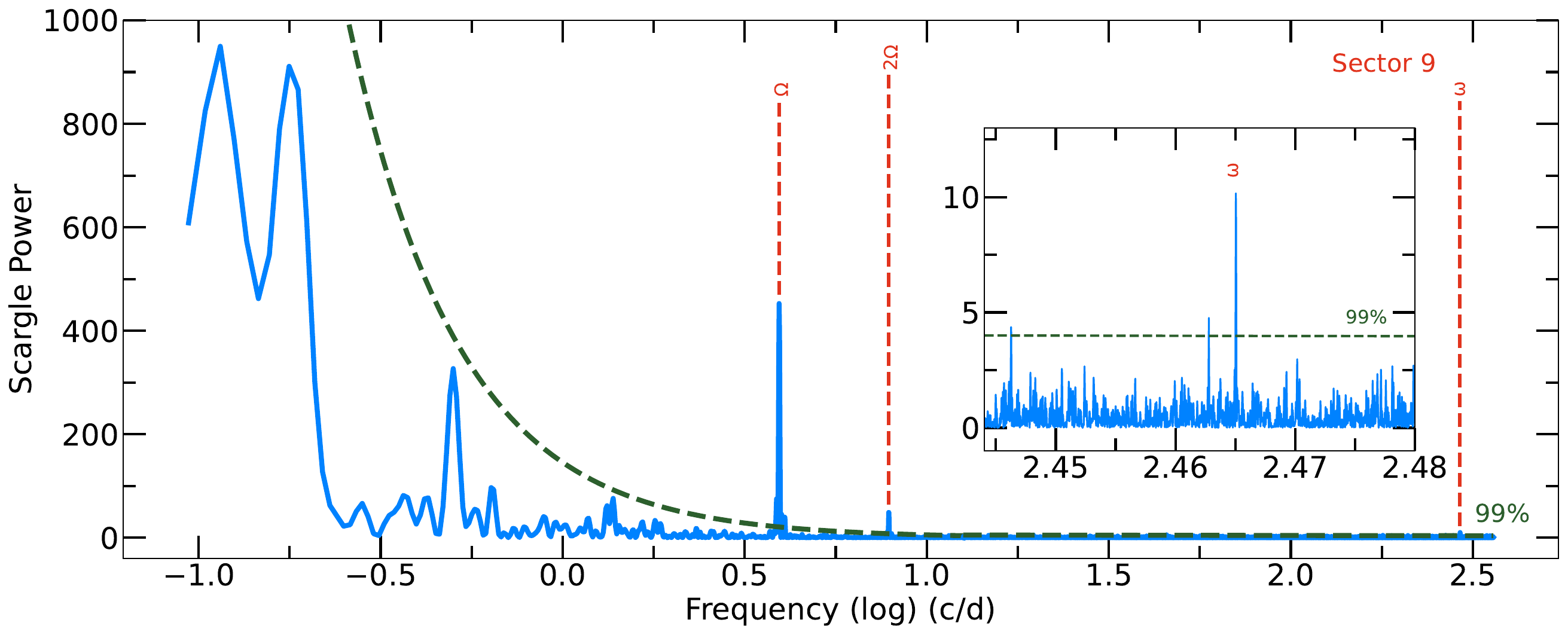}\label{fig:ps_J09}}
    \caption{(a) \textit{TESS} light curve of approximately two consecutive days of J0958 from the observation of sector 9. (b) \textit {TESS} power spectra of J0958 for sector 9, where the identified frequencies are marked. The green horizontal dashed lines represent the 99\% confidence level. The $\omega$ frequency is shown in the zoomed-in plot.}
    \label{fig:lc_ps_J09}
\end{figure}

\subsection{J045707}
\subsubsection{Light curves and power spectra}
Figure \ref{fig:lc_J04} displays the light curve for J045707 for approximately one orbital cycle, whereas Figure \ref{fig:ps_J04} shows the LS power spectrum obtained of the entire light curve. The significance level was calculated following the same methodology which was applied to the previous source. Three dominant frequencies were clearly identified above the 99\% confidence limit. The most dominant peak in the power spectrum was found at 1218.7 s, which is the \psp. The second most dominant peak was found at 6.09 h, which corresponds to the \porb of the system. Using values of the \psp and the \porb, we have derived the beat period (\pbe) of the system as $\sim$ 1290.4 s, which is clearly evident in the power spectra with the period of 1290.5 s. All the significant periods derived from power spectra are given in Table \ref{tab:ps}.

\begin{figure}[b]
    \centering
    \subfigure[Orbital phase folded light curve]{\includegraphics[width=\columnwidth]{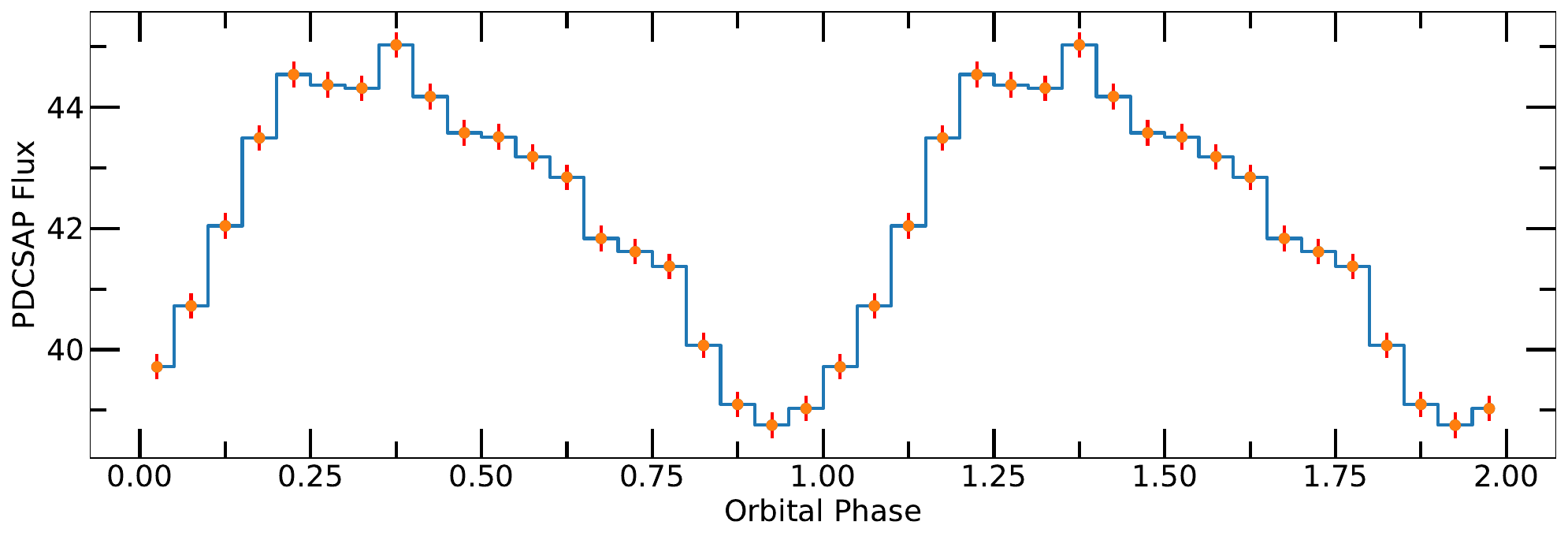}\label{fig:orbfold_J09}} 
    \subfigure[Spin phase folded light curve]{\includegraphics[width=\columnwidth]{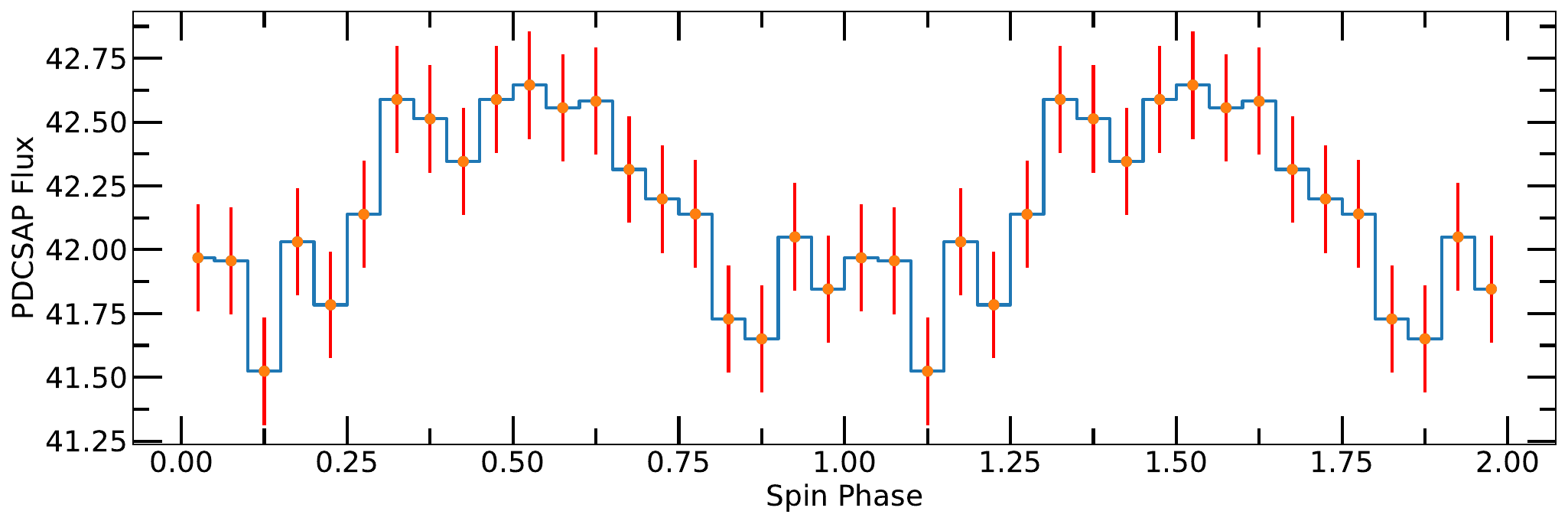}\label{fig:spfold_J09}} 
    \caption{(a) Orbital and (b) Spin phase folded light curve of J0958 with a binning of 20.}
\end{figure}

\subsubsection{Phase folded light curves}
We have folded the whole sector's light curve using the ephemeris BJD = 2459991.6626 + 0.253875 E. Figure \ref{fig:orbfold_J04} shows the orbital phase folded light curve of J045707, which confirms the derived orbital period. We have also folded the data using the previously suggested orbital periods of 4.8 and 6.2 h by \cite{Thorstensen2013}, but no regular pattern of phased light curve was found. We have also folded the entire sector's light curve with the \psp, which is shown in Figure \ref{fig:spfold_J04}. The spin-phased light curve exhibits a single-peaked pulse profile. To see the evolution of the spin pulse profile, we have folded each day's light curves over our derived \psp with the first observation time (BJD = 2459911.44) as a reference epoch. The day-wise evolution of the spin pulse profile is shown in Figure \ref{fig:dspfold_J04}. This also shows only one peak throughout the observations.

\subsection{J0958}
\subsubsection{Light curves and power spectra}
J0958 was observed in only one sector of the \textit{TESS} observations and the light curve of approximately two consecutive days is shown in Figure \ref{fig:lc_J09}. Similar to the previous two sources, the LS periodogram was performed to find the underlying periodicities. The power spectra is shown in Figure \ref{fig:ps_J09}. We have predominantly obtained three frequencies at 3.93, 7.85, and 291.77 c/d which were above the 99\% confidence limit. The most dominant frequency at 3.93 c/d corresponds to the period of  6.11 h, which we identify as \porb of the system. The subsequent dominant frequency corresponds to the harmonic of the orbital period. A peak was also found at 296.12 s, similar to that obtained in earlier studies as  \psp of the WD. This is shown in the zoomed-in plot of Figure \ref{fig:ps_J09}. The orbital and spin periods are listed in Table \ref{tab:ps}.

\subsubsection{Phase folded light curves}
We have also folded the entire light curve with a period of 6.11 h to confirm the presence of  \porb. The phase folded light curve is shown in Figure \ref{fig:orbfold_J09}. The nature of the folded light curve confirms that this period is indeed the true \porb of the system. Similarly, the entire sector's lightcurve was also folded across the obtained \psp and the corresponding spin phase folded light curve is shown in Figure \ref{fig:spfold_J09}.  The spin-folded light curve exhibits two peaks of unequal intensity, with the secondary peak only marginally exceeding the minimum flux. We have also performed day-wise evolution of spin pulse,  but we could not see any periodic pattern in the phased light curve.

\begin{figure}
    \centering
    \subfigure[Representative light curve]{\includegraphics[width=\columnwidth]{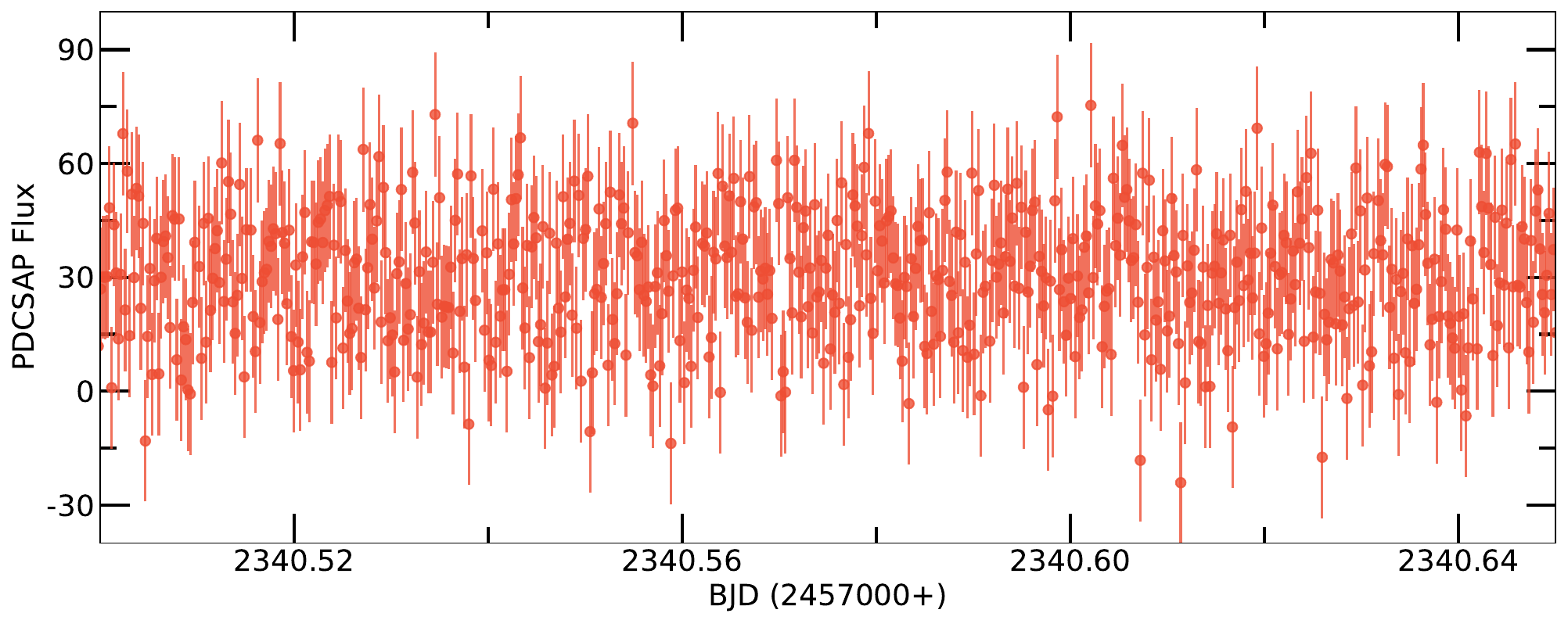}\label{fig:lc_V842}}
    \subfigure[Power spectra]{\includegraphics[width=\columnwidth]{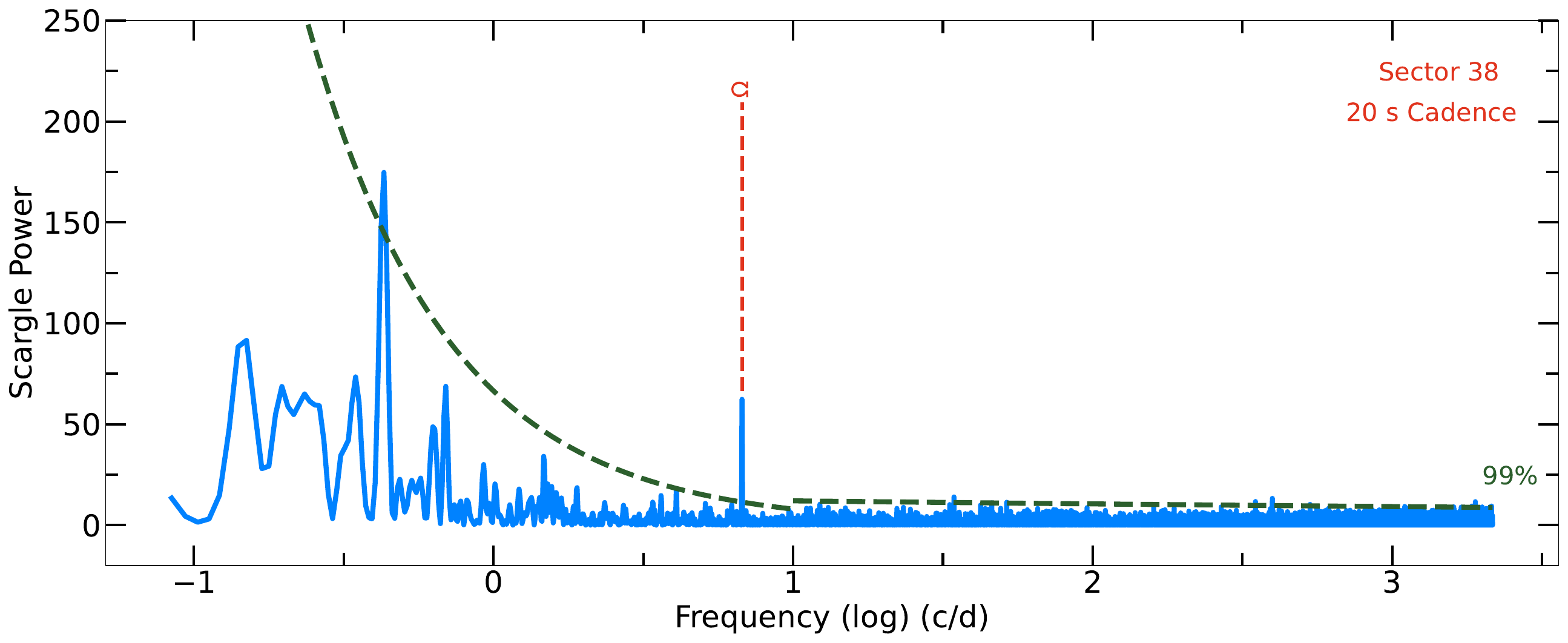}\label{fig:ps_V842}}
    \subfigure[Folded light curve with period of 3.555 h.]{\includegraphics[width=\columnwidth]{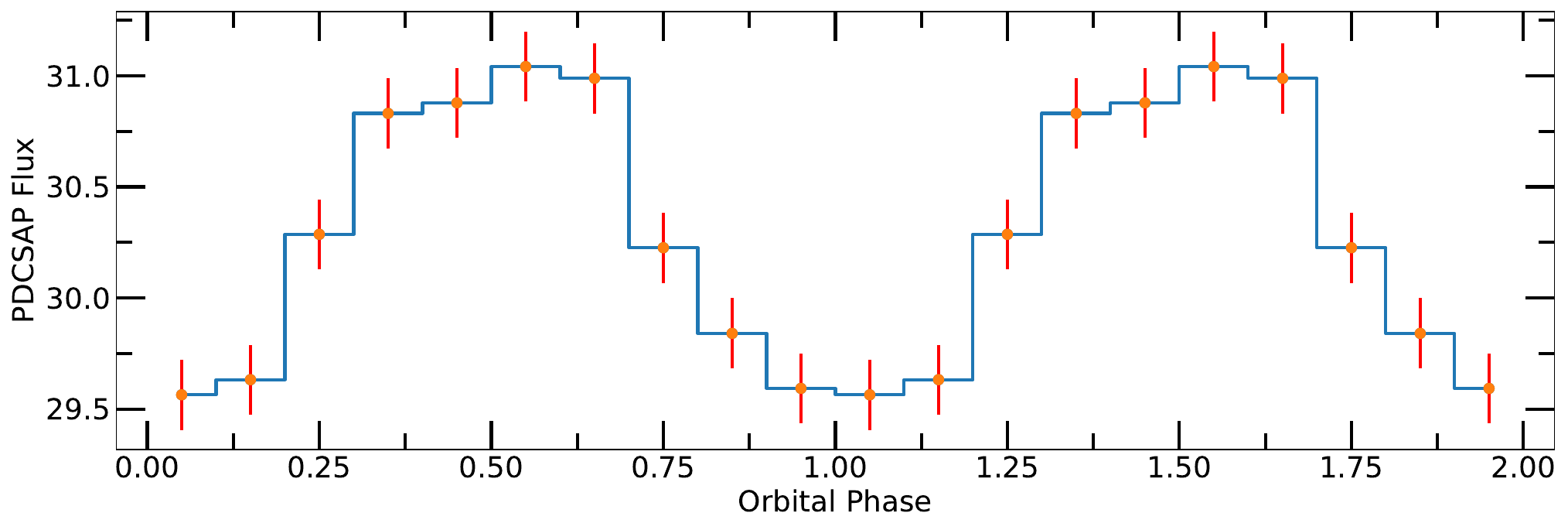}\label{fig:orbfold_V842}}
    \subfigure[Folded light curve with period of 56.5 s.]{\includegraphics[width=\columnwidth]{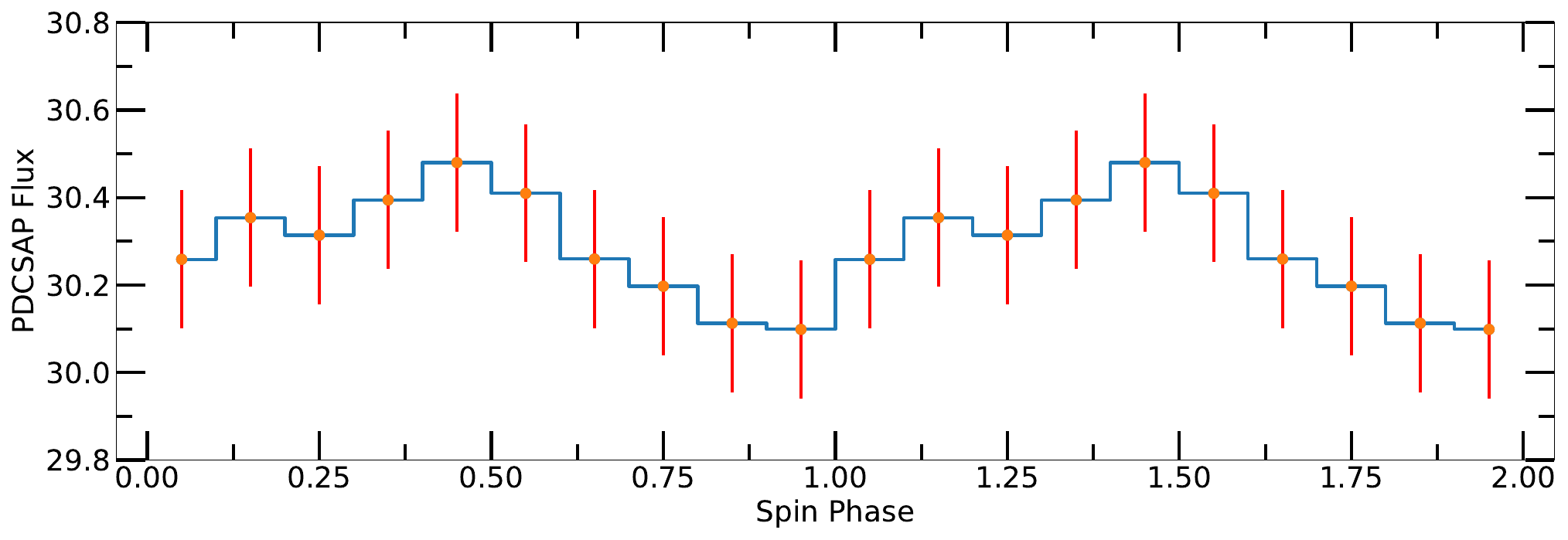}\label{fig:spfold_V842}}
    \caption{(a) \textit{TESS} light curve of approximately 4 h of V842 Cen from the observation of 20 s cadence data of sector 38. (b) \textit{TESS} power spectra of V842 Cen for 20 s cadence data of sector 38, where the identified frequency is marked in red dashed line. The green horizontal dashed line represents the 99$\%$ confidence level. (c) and (d) Folded light curves of V842 Cen over the period of 3.555 h and 56.5 s with a phase bin of 0.1.}
\end{figure}

\subsection{V842 Cen}
\subsubsection{Light curves and power spectra}
V842 Cen was observed in sector 38 during the \textit{TESS} observations. Both 120 and 20 s cadence data were available, and the light curve of approximately 4 h for the 20 s cadence data is shown in Figure \ref{fig:lc_V842}. LS periodogram analysis was performed on both 120 and 20 s cadence datasets, and the power spectra corresponding to the 20 s cadence data is shown in Figure \ref{fig:ps_V842}. The frequencies 6.751 and 0.431 c/d with corresponding periods of 3.555 h and 2.32 d, respectively, were found above the  99\% of the confidence level in the power spectra of 20 s cadence data. However, only the period 3.555 h was found to be consistent in the power spectra of 20 and 120 s cadences' light curves. Therefore, we have not considered the lower frequency for further analysis.

The earlier reported \psp of 56.825/56.5/113.6 s lay beyond the Nyquist limit of 120 s cadence data but was present inside the Nyquist limit for the 20 s cadence data. However, we have not observed any peak corresponding to these frequencies.

\subsubsection{Phase folded light curves}
We have folded the 20 s cadence light curve corresponding to the obtained period of 3.555 h from the power spectral analysis, and it is shown in Figure \ref{fig:orbfold_V842}. A similar nature of the plot was also obtained for the 120 s cadence data, so this confirms the presence of 3.555 h periodicity; thus, it could be the probable  \porb of V842 Cen. We have also folded the light curve with earlier inferred \porb by \cite{Woudt2009}, but we did not find any periodic pattern over these periods. We searched for spin modulations in the \textit{TESS} light curve by folding the data at three previously reported periods: 56.5 s, 56.825 s, and 113.6 s. While folding at 56.825 s and 113.6 s revealed no discernible periodic patterns, a clear periodic signal was seen when using the 56.5 s period. Figure \ref{fig:spfold_V842} shows the folded light curve over the period of 56.5 s. For the folding of the light curve,  we have used the ephemeris provided by \cite{Warner2015}.

\section{Discussion}
\label{sec:dis}
We have carried out a detailed timing analysis of two IPs and two IP candidates using the long time-series high-cadence optical photometric \textit{TESS} observations.  
\subsection{V1460 Her}
The orbital period of V1460 Her is derived to be $4.98845829 \pm 0.00000009$ h, which is more precise than the earlier studies. Based on the O-C analysis, we observed that the orbital period is decreasing at a rate of 1.23 seconds per century. This gradual decrease in the orbital period could be due to the angular momentum loss, which reduces the separation between the primary and secondary, resulting in a decrease in the orbital period. As it has a period above the period gap ($\geq$ 3 h), hence the dominant angular momentum loss mechanism is through magnetic braking \citep[e.g. see][]{Rappaport1983,Schreiber2010}. 

The presence of double-humped light curve features and the detection of the dominant power at $2\Omega$ in V1460 Her suggest that these variations result from the ellipsoidal modulation of the secondary \citep{Warner1995}. The two maxima in the light curves are the result of maximum visibility of the deformed sides of the secondary at orbital phases $\sim$0.75 and $\sim$0.25. At the deeper minimum phase during quiescence, the most probable cause of obscuration between the observer and the combined light of the secondary is a faint accretion disc. This deeper minimum phase filled with the emission during the outburst and the flat out-of-eclipse feature suggests that the accretion disc dominates the light, and the contribution from the secondary star becomes negligible or less visible as compared to quiescence. The double-peaked light curve feature was also found by \cite{Zola2017} and \cite{Kjurkchieva2017}. Throughout the entire duration of the \textit{TESS} observations, the light curve is found to be fairly stable, suggesting that the system is presently in a quiescent state. 

Further, we carefully inspected both maxima of phased light curves and found that in some epochs, the maxima at 0.25 phase are less bright than the ones at 0.75 phase. In some epochs, a reverse trend is seen. This asymmetrical nature of the maxima can be explained by the presence of magnetic activities in the secondary, often known as the O'Connell effect \citep{O'connell1951}. A similar result was also noticed by \cite{Qian2017}.

\subsection{J045707}
The periodogram analysis of J045707 clearly reveals a dominant period of $6.093 \pm 0.002$ h, which we suggested as the probable \porb of the system. \cite{Thorstensen2013} suggested possible \porb of either 4.8 or 6.2 h for this system. We have shown that the period 6.093 h folds the data better than other orbital periods suggested in earlier studies.  Thus, we confirm that the \porb of this system is 6.093 h, which agrees with the period obtained by \cite{2024AJ....168..121B}. We have found the similar \psp of WD as $1218.7 \pm 0.5$ s to that found by \cite{Thorstensen2013}. Considering our derived values of \porb and \psp, the \pbe is estimated to be $\sim$ 1290 s, which is also present in the power spectrum. This is the first time that a significant \be frequency is detected for J045707 due to the better time-cadence and longer baseline of the \textit{TESS}. 

The detection of a beat modulation in J045707 cannot be attributed to the amplitude modulation of the \spin at the \orb. If it were caused by the orbital modulation at the \spin frequency, both \be and $\omega+\Omega$ frequency should be present in the power spectrum with equal power (\citealt{Warner1986}), which is contrary to what is observed in the present dataset. Thus, the detection of \be seems intrinsic.  

J045707 has a slow spinning WD (1218.7 s), and the spin-to-orbital period ratio is determined to be 0.056. Such systems are expected to be disc-fed systems, but the presence of \be frequency also shows the possibility of it being a disc-overflow system. However, an X-ray detection of the \be frequency is required to confirm the accretion geometry as \be in the optical band may also arise due to a spin-varying X-ray beam illuminating structures, such as the secondary star or the hotspot, that are fixed in the orbital frame.  In the case of stream-fed accretion, \cite{Ferrario1999} suggested that a significant power in \be frequency can be detected in the optical band, if the up (above orbital plane) and down (below orbital plane) field symmetry is broken. As J045707 is a relatively slow rotator, it must have a strong magnetic field \citep{Allan1996}.

\subsection{J0958}
For the first time, we have found clear evidence of \porb of 6.11 h. 
\cite{Bernardini2017} proposed that this source exhibits long-term X-ray variability on a time scale of approximately 8 h. However, our results indicate a shorter \porb. Additionally, we explored a potential variability of around 2.2 h, as mentioned by \cite{Bernardini2017}, but found no evidence for this in the \textit{TESS} dataset. We have also obtained a \psp of 296.12 s, which is almost consistent with the results obtained by \cite{Bernardini2017}. 

From the obtained values of \porb and \psp, the \pbe of the system is estimated to be 300.16 s, but we did not get any peak around this value in the power spectra. This suggests the system to be a disc-fed accretor. The value of the spin-to-orbital period is derived to be 0.013( $\leq 0.1$ ). This small value also supports the favour of disc-fed accretion. Contrary to J045707, the source J0958 is a fast rotator (296.12 s). Hence, it must have a weak magnetic field.

\subsection{V842 Cen}
We have detected a period of  3.555 h in this system. Earlier \cite{Woudt2009} had inferred the \porb to be 3.94 h from the values obtained from the \psp and other sidebands, which we could not detect from the \textit{TESS} observations. However,  our derived period is consistent with that derived orbital period by \cite{Luna2012} using X-ray data. Again, the inferred \porb of 3.94 h may not be the orbital period of the system,  when considering the \psp of WD as 113.6 s as proposed by \cite{Warner2015}. Thus, a period of 3.555 h  appears to be an orbital period of V842 Cen. Contrary to earlier findings, the \textit{TESS} power spectrum showed no periodic signals at previously identified spin frequencies. Also, the past investigation of \textit{ROSAT, XMM-Newton}, and \textit{Swift} X-ray observations yielded no detection of pulsed emission (\psp). The folded light curve, exhibiting a 56.5 s periodicity, suggests this as a likely spin period for V842 Cen.   However, further X-ray and optical observations are required to constrain these periodicities. 

\section{Conclusions}
The long-term photometric \textit{TESS} observations of V1460 Her, J045707, J0958, and V842 Cen lead us to the following conclusions. 
\begin{itemize}
    \item Using the extensive \textit{TESS} observations, we have derived a more precise orbital period of V1460 Her as 4.98845829 $\pm$ 0.00000009 h. The long-term photometric data revealed a decreasing trend in the orbital period of V1460 Her.  
    
    \item We have confirmed the orbital period and, for the first time, detected a beat period of $1290.6 \pm 0.5$ s in the source J045707. While the strong spin frequency suggests that disc-fed accretion is the primary mode of accretion, the detection of the beat frequency indicates that some material may be accreting directly onto the white dwarf along its magnetic field lines.

   \item    The orbital period of 6.11 h for the source J0958 has been determined for the first time. The present analysis indicates that  J0958 is a disc-fed accreting system.

   \item Present analysis suggests an orbital period of 3.555 h for the source V842 Cen. The \textit{TESS} data also indicates the presence of a previously reported spin periodicity of 56.5 s. A lack of clear identification of a spin period for V842 Cen in both the current \textit{TESS} observations and previous X-ray data continues to cast uncertainty on its status as an IP. 
    
\end{itemize}
\label{sec:conc}

\begin{acknowledgements}
We thank the referee for reading our manuscript and providing useful comments and suggestions. This paper includes data collected by the TESS mission funded by NASA's Science Mission Directorate. We acknowledge with thanks the variable star observations from the AAVSO International Database contributed by observers worldwide and used in this research. The CRTS and ZTF surveys are supported by the U.S. National Science Foundation under grants AST-0909182/AST-1313422 and  AST-1440341/AST-2034437, respectively. The superWASP data is provided by the WASP consortium and services at the NASA Exoplanet Archive, which is operated by the California Institute of Technology, under contract with the National Aeronautics and Space Administration under the Exoplanet Exploration Program.           
\end{acknowledgements}

\bibliographystyle{aa}
\bibliography{references}

\begin{thebibliography}{49}
\expandafter\ifx\csname natexlab\endcsname\relax\def\natexlab#1{#1}\fi

\bibitem[{{Aizu}(1973)}]{Aizu1973}
{Aizu}, K. 1973, Progress of Theoretical Physics, 49, 1184

\bibitem[{{Allan} {et~al.}(1996){Allan}, {Horne}, {Hellier}, {Mukai}, {Barwig}, {Bennie}, \& {Hilditch}}]{Allan1996}
{Allan}, A., {Horne}, K., {Hellier}, C., {et~al.} 1996, \mnras, 279, 1345

\bibitem[{{Ashley} {et~al.}(2020){Ashley}, {Marsh}, {Breedt}, {G{\"a}nsicke}, {Pala}, {Toloza}, {Chote}, {Thorstensen}, \& {Burleigh}}]{Ashley2020}
{Ashley}, R.~P., {Marsh}, T.~R., {Breedt}, E., {et~al.} 2020, \mnras, 499, 149

\bibitem[{{Bellm} {et~al.}(2019){Bellm}, {Kulkarni}, {Graham}, {Dekany}, {Smith}, {Riddle}, {Masci}, {Helou}, {Prince}, {Adams}, {Barbarino}, {Barlow}, {Bauer}, {Beck}, {Belicki}, {Biswas}, {Blagorodnova}, {Bodewits}, {Bolin}, {Brinnel}, {Brooke}, {Bue}, {Bulla}, {Burruss}, {Cenko}, {Chang}, {Connolly}, {Coughlin}, {Cromer}, {Cunningham}, {De}, {Delacroix}, {Desai}, {Duev}, {Eadie}, {Farnham}, {Feeney}, {Feindt}, {Flynn}, {Franckowiak}, {Frederick}, {Fremling}, {Gal-Yam}, {Gezari}, {Giomi}, {Goldstein}, {Golkhou}, {Goobar}, {Groom}, {Hacopians}, {Hale}, {Henning}, {Ho}, {Hover}, {Howell}, {Hung}, {Huppenkothen}, {Imel}, {Ip}, {Ivezi{\'c}}, {Jackson}, {Jones}, {Juric}, {Kasliwal}, {Kaspi}, {Kaye}, {Kelley}, {Kowalski}, {Kramer}, {Kupfer}, {Landry}, {Laher}, {Lee}, {Lin}, {Lin}, {Lunnan}, {Giomi}, {Mahabal}, {Mao}, {Miller}, {Monkewitz}, {Murphy}, {Ngeow}, {Nordin}, {Nugent}, {Ofek}, {Patterson}, {Penprase}, {Porter}, {Rauch}, {Rebbapragada}, {Reiley}, {Rigault}, {Rodriguez}, {van Roestel}, {Rusholme}, {van
  Santen}, {Schulze}, {Shupe}, {Singer}, {Soumagnac}, {Stein}, {Surace}, {Sollerman}, {Szkody}, {Taddia}, {Terek}, {Van Sistine}, {van Velzen}, {Vestrand}, {Walters}, {Ward}, {Ye}, {Yu}, {Yan}, \& {Zolkower}}]{Bellm2019}
{Bellm}, E.~C., {Kulkarni}, S.~R., {Graham}, M.~J., {et~al.} 2019, \pasp, 131, 018002

\bibitem[{{Bernardini} {et~al.}(2015){Bernardini}, {de Martino}, {Mukai}, {Israel}, {Falanga}, {Ramsay}, \& {Masetti}}]{Bernardini2015}
{Bernardini}, F., {de Martino}, D., {Mukai}, K., {et~al.} 2015, \mnras, 453, 3100

\bibitem[{{Bernardini} {et~al.}(2017){Bernardini}, {de Martino}, {Mukai}, {Russell}, {Falanga}, {Masetti}, {Ferrigno}, \& {Israel}}]{Bernardini2017}
{Bernardini}, F., {de Martino}, D., {Mukai}, K., {et~al.} 2017, \mnras, 470, 4815

\bibitem[{{Bruch}(2024)}]{2024AJ....168..121B}
{Bruch}, A. 2024, \aj, 168, 121

\bibitem[{{Cropper}(1990)}]{Cropper1990}
{Cropper}, M. 1990, \ssr, 54, 195

\bibitem[{{Drake} {et~al.}(2009){Drake}, {Djorgovski}, {Mahabal}, {Beshore}, {Larson}, {Graham}, {Williams}, {Christensen}, {Catelan}, {Boattini}, {Gibbs}, {Hill}, \& {Kowalski}}]{Drake2009}
{Drake}, A.~J., {Djorgovski}, S.~G., {Mahabal}, A., {et~al.} 2009, \apj, 696, 870

\bibitem[{{Drake} {et~al.}(2016){Drake}, {Djorgovski}, {Mahabal}, {Graham}, {Stern}, {Catelan}, {Christensen}, \& {Larson}}]{Drake2016}
{Drake}, A.~J., {Djorgovski}, S.~G., {Mahabal}, A.~A., {et~al.} 2016, The Astronomer's Telegram, 9319, 1

\bibitem[{{Ferrario} \& {Wickramasinghe}(1999)}]{Ferrario1999}
{Ferrario}, L. \& {Wickramasinghe}, D.~T. 1999, \mnras, 309, 517

\bibitem[{{Hart} {et~al.}(2023){Hart}, {Shappee}, {Hey}, {Kochanek}, {Stanek}, {Lim}, {Dobbs}, {Tucker}, {Jayasinghe}, {Beacom}, {Boright}, {Holoien}, {Ong}, {Prieto}, {Thompson}, \& {Will}}]{Hart2023}
{Hart}, K., {Shappee}, B.~J., {Hey}, D., {et~al.} 2023, arXiv e-prints, arXiv:2304.03791

\bibitem[{{Hellier}(1991)}]{Hellier1991}
{Hellier}, C. 1991, \mnras, 251, 693

\bibitem[{{Hellier}(1993)}]{Hellier1993}
{Hellier}, C. 1993, \mnras, 265, L35

\bibitem[{{Hellier} {et~al.}(1989{\natexlab{a}}){Hellier}, {Mason}, \& {Cropper}}]{Hellier1989a}
{Hellier}, C., {Mason}, K.~O., \& {Cropper}, M. 1989{\natexlab{a}}, \mnras, 237, 39P

\bibitem[{{Hellier} {et~al.}(1989{\natexlab{b}}){Hellier}, {Mason}, {Smale}, {Corbet}, {O'Donoghue}, {Barrett}, \& {Warner}}]{Hellier1989b}
{Hellier}, C., {Mason}, K.~O., {Smale}, A.~P., {et~al.} 1989{\natexlab{b}}, \mnras, 238, 1107

\bibitem[{{K{\'a}lm{\'a}n} {et~al.}(2025){K{\'a}lm{\'a}n}, {Csizmadia}, {P{\'a}l}, \& {Szab{\'o}}}]{Kalman2025}
{K{\'a}lm{\'a}n}, S., {Csizmadia}, S., {P{\'a}l}, A., \& {Szab{\'o}}, G.~M. 2025, Research Notes of the American Astronomical Society, 9, 33

\bibitem[{{Kaplan} {et~al.}(2006){Kaplan}, {Gaensler}, {Kulkarni}, \& {Slane}}]{Kaplan2006}
{Kaplan}, D.~L., {Gaensler}, B.~M., {Kulkarni}, S.~R., \& {Slane}, P.~O. 2006, \apjs, 163, 344

\bibitem[{{Kjurkchieva} {et~al.}(2017){Kjurkchieva}, {Popov}, {Vasileva}, \& {Petrov}}]{Kjurkchieva2017}
{Kjurkchieva}, D.~P., {Popov}, V.~A., {Vasileva}, D.~L., \& {Petrov}, N.~I. 2017, \na, 52, 8

\bibitem[{{Lohr} {et~al.}(2013){Lohr}, {Norton}, {Kolb}, {Maxted}, {Todd}, \& {West}}]{Lohr2013}
{Lohr}, M.~E., {Norton}, A.~J., {Kolb}, U.~C., {et~al.} 2013, \aap, 549, A86

\bibitem[{{Lomb}(1976)}]{Lomb1976}
{Lomb}, N.~R. 1976, \apss, 39, 447

\bibitem[{{Luna} {et~al.}(2012){Luna}, {Diaz}, {Brickhouse}, \& {Moraes}}]{Luna2012}
{Luna}, G.~J.~M., {Diaz}, M.~P., {Brickhouse}, N.~S., \& {Moraes}, M. 2012, \mnras, 423, L75

\bibitem[{{Masetti} {et~al.}(2012){Masetti}, {Parisi}, {Jim{\'e}nez-Bail{\'o}n}, {Palazzi}, {Chavushyan}, {Bassani}, {Bazzano}, {Bird}, {Dean}, {Galaz}, {Landi}, {Malizia}, {Minniti}, {Morelli}, {Schiavone}, {Stephen}, \& {Ubertini}}]{Masetti2012}
{Masetti}, N., {Parisi}, P., {Jim{\'e}nez-Bail{\'o}n}, E., {et~al.} 2012, \aap, 538, A123

\bibitem[{{Masetti} {et~al.}(2010){Masetti}, {Parisi}, {Palazzi}, {Jim{\'e}nez-Bail{\'o}n}, {Chavushyan}, {Bassani}, {Bazzano}, {Bird}, {Dean}, {Charles}, {Galaz}, {Landi}, {Malizia}, {Mason}, {McBride}, {Minniti}, {Morelli}, {Schiavone}, {Stephen}, \& {Ubertini}}]{Masetti2010}
{Masetti}, N., {Parisi}, P., {Palazzi}, E., {et~al.} 2010, \aap, 519, A96

\bibitem[{{Masetti} {et~al.}(2013){Masetti}, {Parisi}, {Palazzi}, {Jim{\'e}nez-Bail{\'o}n}, {Chavushyan}, {McBride}, {Rojas}, {Steward}, {Bassani}, {Bazzano}, {Bird}, {Charles}, {Galaz}, {Landi}, {Malizia}, {Mason}, {Minniti}, {Morelli}, {Schiavone}, {Stephen}, \& {Ubertini}}]{Masetti2013}
{Masetti}, N., {Parisi}, P., {Palazzi}, E., {et~al.} 2013, \aap, 556, A120

\bibitem[{{McNaught} \& {Mattei}(1986)}]{McNaught1986}
{McNaught}, R.~H. \& {Mattei}, J.~A. 1986, \iaucirc, 4284, 4

\bibitem[{{O'Connell}(1951)}]{O'connell1951}
{O'Connell}, D.~J.~K. 1951, Publications of the Riverview College Observatory, 2, 85

\bibitem[{{Patterson}(1994)}]{Patterson1994}
{Patterson}, J. 1994, \pasp, 106, 209

\bibitem[{{Pelisoli} {et~al.}(2021){Pelisoli}, {Marsh}, {Ashley}, {Hakala}, {Aungwerojwit}, {Burdge}, {Breedt}, {Brown}, {Chanthorn}, {Dhillon}, {Dyer}, {Green}, {Kerry}, {Littlefair}, {Parsons}, {Sahman}, {Wild}, \& {Yotthanathong}}]{Pelisoli2021}
{Pelisoli}, I., {Marsh}, T.~R., {Ashley}, R.~P., {et~al.} 2021, \mnras, 507, 6132

\bibitem[{{Pollacco} {et~al.}(2006){Pollacco}, {Skillen}, {Collier Cameron}, {Christian}, {Hellier}, {Irwin}, {Lister}, {Street}, {West}, {Anderson}, {Clarkson}, {Deeg}, {Enoch}, {Evans}, {Fitzsimmons}, {Haswell}, {Hodgkin}, {Horne}, {Kane}, {Keenan}, {Maxted}, {Norton}, {Osborne}, {Parley}, {Ryans}, {Smalley}, {Wheatley}, \& {Wilson}}]{Pollocco2006}
{Pollacco}, D.~L., {Skillen}, I., {Collier Cameron}, A., {et~al.} 2006, {The WASP Project and the SuperWASP Cameras}

\bibitem[{{Qian} {et~al.}(2017){Qian}, {Han}, {Zhang}, {Zejda}, {Michel}, {Zhu}, {Zhao}, {Liao}, {Tian}, \& {Wang}}]{Qian2017}
{Qian}, S.~B., {Han}, Z.~T., {Zhang}, B., {et~al.} 2017, \apj, 848, 131

\bibitem[{{Rappaport} {et~al.}(1983){Rappaport}, {Verbunt}, \& {Joss}}]{Rappaport1983}
{Rappaport}, S., {Verbunt}, F., \& {Joss}, P.~C. 1983, \apj, 275, 713

\bibitem[{{Ricker} {et~al.}(2015){Ricker}, {Winn}, {Vanderspek}, {Latham}, {Bakos}, {Bean}, {Berta-Thompson}, {Brown}, {Buchhave}, {Butler}, {Butler}, {Chaplin}, {Charbonneau}, {Christensen-Dalsgaard}, {Clampin}, {Deming}, {Doty}, {De Lee}, {Dressing}, {Dunham}, {Endl}, {Fressin}, {Ge}, {Henning}, {Holman}, {Howard}, {Ida}, {Jenkins}, {Jernigan}, {Johnson}, {Kaltenegger}, {Kawai}, {Kjeldsen}, {Laughlin}, {Levine}, {Lin}, {Lissauer}, {MacQueen}, {Marcy}, {McCullough}, {Morton}, {Narita}, {Paegert}, {Palle}, {Pepe}, {Pepper}, {Quirrenbach}, {Rinehart}, {Sasselov}, {Sato}, {Seager}, {Sozzetti}, {Stassun}, {Sullivan}, {Szentgyorgyi}, {Torres}, {Udry}, \& {Villasenor}}]{Ricker2015}
{Ricker}, G.~R., {Winn}, J.~N., {Vanderspek}, R., {et~al.} 2015, Journal of Astronomical Telescopes, Instruments, and Systems, 1, 014003

\bibitem[{{Rosen} {et~al.}(1988){Rosen}, {Mason}, \& {Cordova}}]{Rosen1988}
{Rosen}, S.~R., {Mason}, K.~O., \& {Cordova}, F.~A. 1988, \mnras, 231, 549

\bibitem[{{Scargle}(1982)}]{Scargle1982}
{Scargle}, J.~D. 1982, \apj, 263, 835

\bibitem[{{Scaringi} {et~al.}(2016){Scaringi}, {Mason}, {Van Winckel}, \& {Escorza}}]{Scaringi2016}
{Scaringi}, S., {Mason}, E., {Van Winckel}, H., \& {Escorza}, A. 2016, The Astronomer's Telegram, 9122, 1

\bibitem[{{Schreiber} {et~al.}(2010){Schreiber}, {G{\"a}nsicke}, {Rebassa-Mansergas}, {Nebot Gomez-Moran}, {Southworth}, {Schwope}, {M{\"u}ller}, {Papadaki}, {Pyrzas}, {Rabitz}, {Rodr{\'\i}guez-Gil}, {Schmidtobreick}, {Schwarz}, {Tappert}, {Toloza}, {Vogel}, \& {Zorotovic}}]{Schreiber2010}
{Schreiber}, M.~R., {G{\"a}nsicke}, B.~T., {Rebassa-Mansergas}, A., {et~al.} 2010, \aap, 513, L7

\bibitem[{{Sekiguchi} {et~al.}(1989){Sekiguchi}, {Feast}, {Fairall}, \& {Winkler}}]{Sekiguchi1989}
{Sekiguchi}, K., {Feast}, M.~W., {Fairall}, A.~P., \& {Winkler}, H. 1989, \mnras, 241, 311

\bibitem[{{Shappee} {et~al.}(2014){Shappee}, {Prieto}, {Grupe}, {Kochanek}, {Stanek}, {De Rosa}, {Mathur}, {Zu}, {Peterson}, {Pogge}, {Komossa}, {Im}, {Jencson}, {Holoien}, {Basu}, {Beacom}, {Szczygie{\l}}, {Brimacombe}, {Adams}, {Campillay}, {Choi}, {Contreras}, {Dietrich}, {Dubberley}, {Elphick}, {Foale}, {Giustini}, {Gonzalez}, {Hawkins}, {Howell}, {Hsiao}, {Koss}, {Leighly}, {Morrell}, {Mudd}, {Mullins}, {Nugent}, {Parrent}, {Phillips}, {Pojmanski}, {Rosing}, {Ross}, {Sand}, {Terndrup}, {Valenti}, {Walker}, \& {Yoon}}]{Shappee2014}
{Shappee}, B.~J., {Prieto}, J.~L., {Grupe}, D., {et~al.} 2014, \apj, 788, 48

\bibitem[{{Sion} {et~al.}(2013){Sion}, {Szkody}, {Mukadam}, {Warner}, {Woudt}, {Walter}, {Henden}, \& {Godon}}]{Sion2013}
{Sion}, E.~M., {Szkody}, P., {Mukadam}, A., {et~al.} 2013, \apj, 772, 116

\bibitem[{{Suleimanov} {et~al.}(2019){Suleimanov}, {Doroshenko}, \& {Werner}}]{Suleimanov2019}
{Suleimanov}, V.~F., {Doroshenko}, V., \& {Werner}, K. 2019, \mnras, 482, 3622

\bibitem[{{Thorstensen} \& {Halpern}(2013)}]{Thorstensen2013}
{Thorstensen}, J.~R. \& {Halpern}, J. 2013, \aj, 146, 107

\bibitem[{{Vaughan}(2005)}]{Vaughan2005}
{Vaughan}, S. 2005, \aap, 431, 391

\bibitem[{{Warner}(1986)}]{Warner1986}
{Warner}, B. 1986, \mnras, 219, 751

\bibitem[{{Warner}(1995)}]{Warner1995}
{Warner}, B. 1995, Cambridge Astrophysics Series, 28

\bibitem[{{Warner} \& {Woudt}(2015)}]{Warner2015}
{Warner}, B. \& {Woudt}, P.~A. 2015, \memsai, 86, 108

\bibitem[{{Woudt} \& {Warner}(2003)}]{Woudt2003}
{Woudt}, P.~A. \& {Warner}, B. 2003, \mnras, 340, 1011

\bibitem[{{Woudt} {et~al.}(2009){Woudt}, {Warner}, {Osborne}, \& {Page}}]{Woudt2009}
{Woudt}, P.~A., {Warner}, B., {Osborne}, J., \& {Page}, K. 2009, \mnras, 395, 2177

\bibitem[{{Zola} {et~al.}(2017){Zola}, {Szkody}, {Ciprini}, {Verrecchia}, {Debski}, {Ogloza}, {Drozdz}, {Reichart}, {Caton}, \& {Hoette}}]{Zola2017}
{Zola}, S., {Szkody}, P., {Ciprini}, S., {et~al.} 2017, \aj, 154, 276

\end{thebibliography}

\end{document}